\documentclass{ar-1col}

\usepackage{hyper ref}
\usepackage{natbib}
\usepackage{amsmath}
\usepackage{amssymb}

\jname{Annual Review of Earth and Planetary Sciences}
\jvol{47}
\jyear{2018}
\doi{10.1146/((please add article doi))}

\begin{document}

\markboth{Christiane Helling}{Exoplanet Clouds}

\title{Exoplanet Clouds}

\author{Christiane Helling
\affil{Centre for Exoplanet Science, University of St Andrews,\\ St Andrews KY16 9SS, UK;\\ email: ch80@st-andrews.ac.uk}}

\begin{abstract}
\begin{minipage}{9cm}
Clouds also form in atmospheres of  planets that orbit other stars than our Sun, in so-called extrasolar planets or exoplanets. Exoplanet atmospheres can be chemically extremely rich. Exoplanet clouds are therefor made of a mix of materials that changes throughout the atmosphere. They affect the atmospheres through element depletion and through absorption and
scattering, hence, they have a profound impact on the atmosphere's energy budget.  While astronomical observations point us to the presence  of extrasolar clouds and make first suggestions on particle sizes and material compositions, we require fundamental and complex modelling work to merge the individual observations into a coherent picture. Part of this is to develop an understanding for cloud formation in non-terrestrial environments.
\end{minipage}
\end{abstract}

\maketitle

\tableofcontents

\section{INTRODUCTION}
By June 2018, 3796 extrasolar planet were known to exist. These planets
are residing in 2840 planetary systems of which 632 contain multiple
planets (Exoplanet.eu). If both mass and radius are known, a mean
density can be derived which provides us a first idea about if the
planet is made of light elements like hydrogen and helium, or if it
contains a large fraction of water, rock and/or iron. The planetary
composition is determined by the planet's formation process and the
material that is accreted onto the planet from the planetary
disk. Planets that form very close to their host star will
predominantly accrete materials that contain large amounts of
Mg/Si/Fe/O, hence, they are termed 'rocky planets'. Planets that form
at the outskirts of the planet forming disks, where temperatures are
low enough that ices of all sorts (H$_2$O ice, CO ice, CH$_4$ ice)
form, will consequently contain carbon, hydrogen and considerable
amounts of oxygen but also large amounts of Mg/Si/Fe as part of their
cores.  The planetary atmosphere forms through gas accretion and,
later, mainly through outgassing (incl. volcanism) of the accreted
disk material. We may therefore expect to find fingerprints of the
planet formation process by studying its atmosphere. However, our
enthusiasm has been challenged by the discovery that many, if not all,
extrasolar planets will be enshrouded by clouds. This realisation will
have a considerable impact on target selection for upcoming space
missions (e.g., James-Webb Space Telescope (JWST) that aims to characterise
exoplanet atmospheres in the infrared spectral range) for which
planets were expected to be entirely cloud-free. Only the hottest
exoplanets that are closest to their host star may expose a clear
dayside and some a lava-like surface.

\section{Exoplanet cloud observations}
Clouds on exoplanets were postulated as the reason for missing
molecular or atomic absorption features expected to be detected in the
optical and near-IR spectra as obtained from from space-
(e.g. Hubble-Space Telescope (HST) which took the first picture of an
exoplanet in the optical spectral range) and ground-based (e.g with
the optical Very Large Telescope composed of four 8.2 m telescopes
that can be combined into an interferometer) observations.  Brown
Dwarfs were the first class of objects discover to have extensive
atmospheric clouds. Brown Dwarfs undergo different formation processes
than planets. They form as stars (hence, ingnite some nuclear fusion
which planets don't) and not as part of a stellar accretion
disk. Clouds have an distinct impact on the atmospheric profiles of
Brown Dwarfs and planets. In order to reflect his, different classes
of Brown Dwarfs were suggested for classification purposes: L dwarfs
(clouds in observable atmosphere) and T dwarfs (clouds affect
atmosphere but are below observable atmosphere). More recently, the Y
dwarfs were added. These are Brown dwarfs, hence  stellar objects,
that are most similar to planets regarding their very low global
temperatures (T$_{\rm eff}<200$K; T$_{\rm eff}$ is the effective
temperature and it measures the total radiation output of an
object). Since their discovery, Brown Dwarfs have been studied as
analogues of extrasolar planets also with respect to their atmospheric
processes. This is supported particularly by the overlapping mass
regimes, but also by the location of very low gravity Brown Dwarfs and
free-floating planets in the colour-magnitude (J-K) diagram
(\citealt{2018ApJ...854..172C}). Brown Dwarfs further exhibit
rotating, banded cloud structure similar to what is seen on Jupiter in
optical wavelengths. Their atmospheric dynamics resembles best the
Neptunian atmosphere (\citealt{2017Sci...357..683A}). Given such
similarity, we expect to find vertically (and horizontally) extended
cloud structures in exoplanets, similar to what has already been
spectroscopically demonstrated for Brown Dwarfs (e.g.,
\citealt{2013ApJ...768..121A}).  The first extrasolar planet for which
an extensive cloud layer was derived from observations is the 1.15
M$_{\rm Jup}$ mass giant gas planet HD\,189733b\footnote{HD refers to
  stars which have been catalogued in the Henry Draper Catalogue. The
  small letter behind the catalogue number (189733) indicates that
  HD\,189733b is a planet orbiting the star HD\,189733.}  orbiting a
K-dwarf star in 2.2 days. \citet{2013MNRAS.432.2917P} present a
transmission spectrum (i.e. observations of the planet transiting in
front of the star at different wavelengths) ranging from the UV to the
near-IR and find that it is dominated by starlight being scattered by
small particles in the planet's upper atmosphere.  A flat optical
exoplanet spectrum without any molecular absorption was also obtained
by persevering HST observations of the super-Earth GJ1214b
(\citealt{2014Natur.505...69K}).  Similarly, the small rocky exoplanet
GJ\,1132b\footnote{GJ refers to stars which are catalogued in the
  Gliese Catalogue of Nearby Stars. The small letter behind the
  catalogue number indicates that it is a planet orbiting the star
  GJ\,1132.} (1.2$R_{\rm Earth}$, 1.6$M_{\rm Earth}$) that orbits a
M-dwarf shows a 0.7$\mu$m-1.04$\mu$m spectrum with no gas absorption
features (\citealt{2018arXiv180507328D}). This could be caused by a
cloud-free atmosphere that is extremely metal rich (10$\times$ solar;
i.e. enriched in elements heavier the hydrogen) which cannot be
H$_2$O/H$_2$-dominated. The only other options are that the rocky
planet has lost its atmosphere, for example through the high-energy
radiation impact from its host star, or that the planet is enshrouded
in vertically and globally extended clouds.

The method that is most widely applied to analyse the chemical
composition of extrasolar planets, at the time of writing, is
transmission spectroscopy.  A transmission spectrum is obtained by
observing the planet transiting in front of a star at different
wavelengths and then measuring the magnitude of the transit depth in the
resulting light curve. It records the opacity of the atmosphere at
different wavelengths. To account for clouds, in the literature such
observations are fitted by presetting the material composition and by
assuming a constant size of cloud particles in the optical, and
alternatively a second size for larger wavelengths. However, exoplanet
cloud particles are neither made of just one material because of their
chemically rich atmospheres, nor do they occur with just one
size. Instead, exoplanet clouds are made of a rich mix of materials in
atmospheres that cover a wide range of temperatures. Most recently,
high-dispersion spectroscopy was developed to analyse the atmospheric
gas also for non-transiting planets (\citealt{2013ApJ...764..182S}).

 Comprehensively understanding exoplanet atmospheres is very much
 linked to the understanding of the cloud formation mechanisms.
 Clouds exert strong feedbacks on an exoplanet atmosphere causing
 visible changes to the observed spectrum:
\begin{itemize}
\item[a)] Cloud particles have a far larger opacity (absorption +
  scattering) than the atmospheric gas, hence, they absorb and re-emit
  more photons. This results in a strong heating of the atmosphere
  below the clouds (greenhouse effect, backwarming) but also in a
  redistribution of the incoming stellar flux through hydrodynamical
  processes. Clouds block our view into the underlying atmosphere such
  that telescopes can only access the atmospheric layers above them.
\item[ b)] During their formation, the cloud particles consume
  elements such that molecules and atoms containing those elements
  (e.g. Si/SiO, Mg/MgOH, Ti/TiO$_2$, O/H$_2$O) will appear less abundant. This effect is best
  observable for less abundant elements like Ti, V or Al showing less
  than expected absorption in e.g. TiO. Any observational element abundance
  determination needs therefore to be conducted with great care as
  only the atmospheric part above an optically thick cloud layer can
  be observed.
\end{itemize}

Astronomical observations provide us with exciting discoveries, but  with only limited information
about any of the exoplanets. We may know the planetary mass and radius, or only
one of the two, the radius ratio planet-to-star at specific
wavelengths (transmission spectrum), the integrated thermal flux or
the albedo.  To merge these pieces of observations into a coherent
picture, chemically and physically consistent models that apply
detailed knowledge of the cloud formation processes, the gas
chemistry, energy transport through radiative and/or convection are
required.  From Earth we know that the diversity of cloud structures
(size, form) is indicative of the local atmospheric conditions and
their history. Thus, the aim of this review is to discuss cloud
formation as a necessary part of a complex modelling approach, which is the
base for our understanding of the plethora of exoplanet atmosphere
observations that will come from space telescopes like CHEOPS\footnote{CHEOPS stands for CHaracterising ExOPlanets Satellite; \url{http://cheops.unibe.ch}}, TESS\footnote{TESS stands for Transiting Exoplanet Survey Satellite; \url{https://www.nasa.gov/tess-transiting-exoplanet-survey-satellite}}, JWST\footnote{JWST stands for James-Webb Space Telescope; \url{https://www.jwst.nasa.gov}},  PLATO\footnote{PLATO stands for PLAnetary Transits and Oscillations of stars and is a space telescope; \url{http://sci.esa.int/plato/}}, ARIEL\footnote{ARIEL stands for Atmospheric Remote-sensing Exoplanet Large-survey and is a space mission; \url{https://ariel-spacemission.eu}}, WFIRST\footnote{Wide Field Infrared Survey Telescope; \url{https://wfirst.gsfc.nasa.gov}}, LUVOIR\footnote{Large UV/Optical/IR Surveyor; \url{https://asd.gsfc.nasa.gov/luvoir/}}  and also from high-precision ground-based instruments like  FORS2\footnote{FORS2 stands for FOcal Reducer/low dispersion Spectrograph 2; \url{https://www.eso.org/sci/facilities/paranal/instruments/fors/overview.html}}, ESPRESSO\footnote{ESPRESSO stands for chelle SPectrograph for Rocky Exoplanets and Stable Spectroscopic Observations; \url{https://www.eso.org/sci/facilities/paranal/instruments/espresso.html}} and CARMENES\footnote{CARMENES stands for Calar Alto high-Resolution search for M dwarfs with Exoearths with Near-infrared and optical {\'E}chelle Spectrographs; \url{https://carmenes.caha.es/ext/instrument/index.html}} at the VLT\footnote{VLT stands for Very Large Telescope; \url{https://www.eso.org/public/teles-instr/paranal-observatory/vlt/}}.

\section{How clouds form on exoplanets}\label{s:how}
The formation of clouds occurs through a sequence of chemical and
physical processes that are determined by the local atmospheric gas
temperature, pressure and chemical composition.  The gas composition
is determined by the number and abundance of elements (H, He, Si, Fe,
Mg, O, C etc.)  available. This initial amount of elements is
determined through stellar (nuclear synthesis) and planetary (pebble
accretion, outgassing, atmosphere loss) evolutionary processes. The
most abundant molecules in a solar-composition, hydrogen-rich
planetary atmosphere are H$_2$, H$_2$O, CO, CH$_4$, N$_2$, NH$_3$
followed by SiO, H$_2$S, Fe(OH)$_2$, Mg(OH)$_2$, and less abundant
molecules like AlO$_2$H, SiS, TiO$_2$ and others
(\citealt{2017arXiv171201010W}) will contribute to the cloud formation
processes.  These atoms, molecules and potentially ions are the
starting point of the cloud formation. Their relative abundances
change if the relative number of the individual elements (O, C, Si,
Mg, Fe, Ca, Al etc.)  changes and the thermodynamic condition of the
gas changes. These processes form a closed feedback circle. Hotter
atmospheres, like the day-side of the rocky planet CoRoT-7b
(\citealt{2017MNRAS.472..447M}) and of ultra-hot Jupiters
(\citealt{2018arXiv180500096P,2018arXiv180500038L}), will contain more
atomic and ionic species, cooler atmospheres will be dominated by
molecules. This effect is well-known from stellar atmospheres
(e.g. \citealt{2008A&A...486..951G}).  For example, the hydrogen ion,
H$^{-}$, is an important opacity in hot Jupiters like WASP-18b because
of their very high days-side temperatures
(\citealt{2018ApJ...855L..30A}). Such high gas temperatures do not
permit cloud formation. Various modelling approaches for considering
clouds as part of atmospheric modelling have been developed and were
compared in \cite{2008MNRAS.391.1854H} and a summarising update is
provided by \cite{2018ApJ...854..172C}. All but one of the approaches
invokes phase-equilibrium (thermal equilibrium). In the following, we
will focus on the cloud formation processes which require the
deviation from phase-equilibrium to enable the actual formation to
occur.

\begin{figure}[h]
\hspace*{.5cm}
\includegraphics[width=16cm]{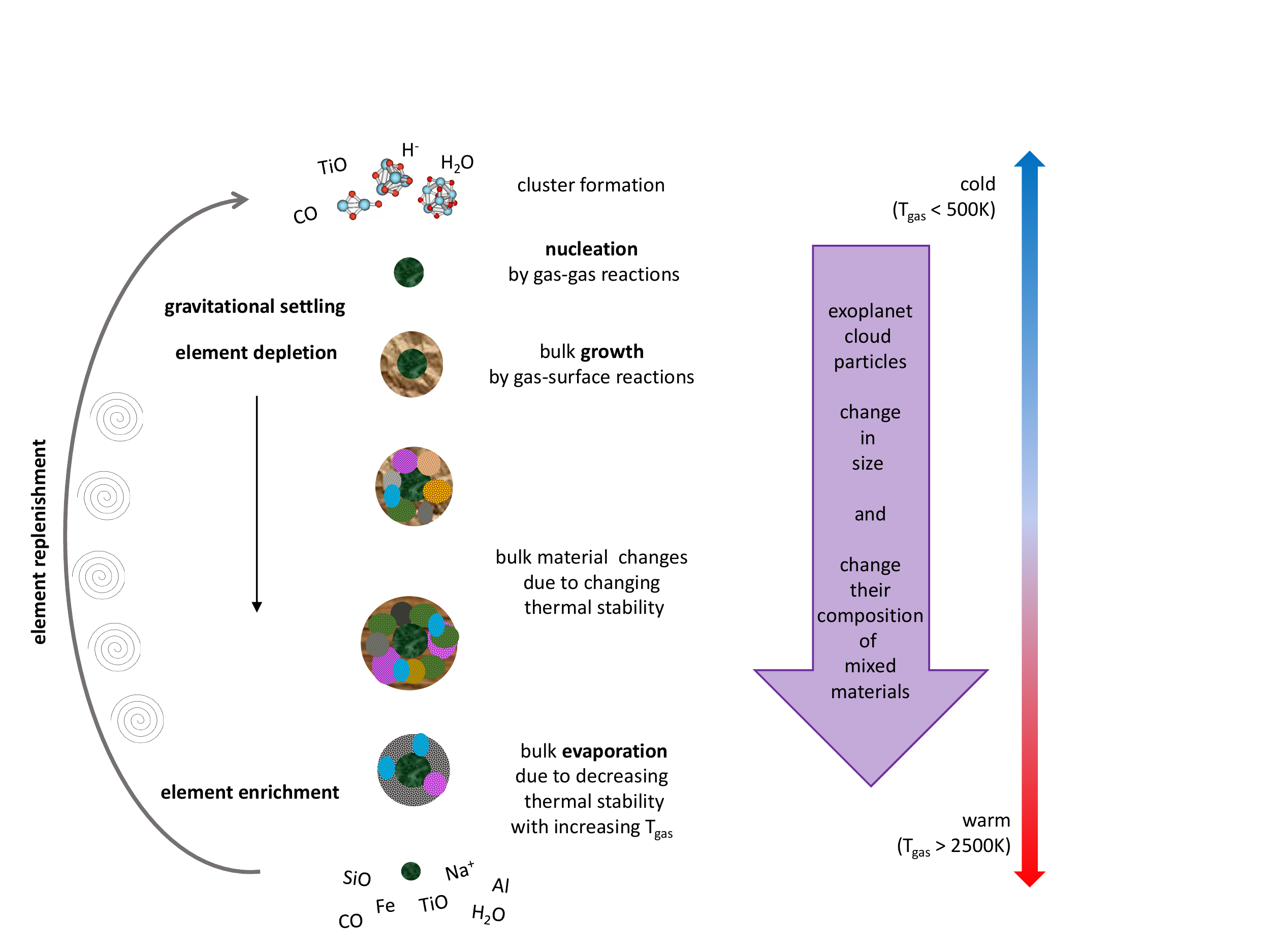}\\*[-0.5cm]
\caption{Cloud formation processes leading to cloud particles made of
  a mix of material that changes with height. Cloud formation in extrasolar planets starts from a chemically very rich gas (including e.g. CO, SiO, CaH, H$_2$O, TiO$_2$, Mg(OH)$_2$). If the temperature is low enough, big molecular structures (so-called clusters) begin to form and grow to condensation nuclei ({\bf nucleation}).  Subsequently, many materials condense onto these seed particles and they {\bf grow}  to macroscopic $\mu$m-sized cloud particles. The cloud particles fall into the atmosphere ({\bf gravitational settling}) where they may change their material composition and sizes completely, until they {\bf evaporate}. The condensation processes {\bf deplete} the atmosphere of elements, and the evaporation process will {\bf enrich} the atmosphere. A stationary cloud forms of an efficient {\bf element replenishment} mechanism is active. Courtesy to David  Gobrecht for the (Al$_2$O$_3$)$_{\rm 1,2,3}$ cluster structure.}
\label{fig1}
\end{figure}

\subsection{Cloud formation processes}

Clouds form through and are influence by a sequence of processes that
are summarised in Figure~\ref{fig1}. These processes are universal and
will therefore occur whenever the local conditions for temperature,
gas density and wind speed are favorable. Cloud formation is seeded by
the emergence of condensation nuclei ({\bf nucleation},
Sect.~\ref{ss:nuc}) onto which all thermally stable materials grow
through gas-surface reactions ({\bf bulk growth},
Sect.~\ref{ss:bulkg}). As the atmosphere is gravitationally bound by
the mass of the planet, the cloud particles will gravitational settle
({\bf gravitational settling}, rain out). As they fall (gravitational
settling, Sect.~\ref{ss:grav}), they will encounter different
temperatures and gas densities which cause the material composition of
the cloud particles to change because of changing thermal
stability. The cloud particles continue to grow until the temperatures
are too high and the materials begin to evaporate. Thermal stability
increases if the local gas pressure increases for a given
temperature. The condensation and evaporation processes change the
local element abundances because participating element are
depleter. This, in turn, chances the species that continue the bulk
growth process. The depletion of oxygen, for example, will lead to an
increase of the local carbon-to-oxygen ratio (C/O). In order for a cloud layer to
persist, element replenishment (Sect.~\ref{ss:grav}) mechanisms in
form of convection and/or diffusion must be active. The vertical,
geometric extension of a cloud is determined by the presence of
condensation nuclei (or their formation) at the upper boundary and by
the thermal stability at the lower boundary.

\subsubsection{Nucleation / seed formation}\label{ss:nuc}

On Earth, most of the cloud particle (aerosols, droplets) form inside
the troposphere where condensation seed particles (cloud condensation
nuclei -- CCN) are available from, for example, sand storms, ocean
spray, fires, combustion and volcano eruptions.  Without the presence
of a condensation seed, spontaneous water condensation from the gas
phase requires a supersaturation of 800\% on Earth which has not been
observed. This statement is well-accepted by scholars in different
disciplines (meteorology, earth and environmental science, astronomy)
but it poses a conundrum. The measurement of the supersaturation in
the Earth atmosphere can not distinguish between the two following
cases: i) The water has condensed onto condensation seeds.  ii) The
thermodynamic conditions are inappropriate for water nucleation
(spontaneous condensation) to occur. The conundrum is that in both
cases the atmospheric gas will be at or near the phase-equilibrium
level and therefore, the gas will not be supersaturated at all (see
gray box).  Both cases lead to the conclusion that water condensation
on Earth can not occur directly from the gas phase. For extrasolar
planets, it is impossible to measure rates for condensation seed
production (like through volcano outbreaks) in situ. It is
therefore necessary for Earth and for extrasolar planets alike to seek
out more fundamental approaches that enable us to predict how many
condensation seeds form from the gas phase to start with. Similar challenges are
known from the formation of clouds in the Earth's mesosphere
(noctilousent/mesospheric clouds), where evaporating meteorites play
their role in forming meteoritic smoke particles onto which H$_2$O ice
condenses (\citealt{2018SSRv..214...23P}), but also from the occurrence
of nucleation bursts in Earth's convective boundary layers
(\citealt{2006ACP.....6.4175H}).  Similar challenges arise for other
solar system planets. For example, Martian mesospheric clouds have
been identified, too, but are predominantly of CO$_2$ ice
(\citealt{2010Icar..209..452M}).

\begin{textbox}[h]
\section{Supersaturation and supercooling}
The concept of {\it thermal stability}  of a material (e.g. TiO$_2$[s], SiO[s]) is defined for a planar,
infinitely extended surface and it occurs if the {\it supersaturation} ratio
S(T)=1. The same concept is applied to non-planar surfaces when considering clouds. Thermal stability represents a state of equilibrium because
the rate of evaporation, $\tau_{\rm ev}$, equals the rate of growth,
$\tau_{\rm gr}$, as shown in the figure below. However, growth and
evaporation are non-equilibrium states as $\tau_{\rm ev}\not=\tau_{\rm
  gr}$. The formation of condensation seeds from the gas phase
requires a considerable supersaturation of the participating gas
because the evaporation rate increases with decreasing cluster size
for any given temperature (\citealt{1996ASPC...96...69G}). The
necessary supersaturation (S$\gg\!\!1$) requires a considerably lower
temperature ({\it supercooling}: T$_{\rm gas}$(S$\gg\!\!1$) $\ll$ T$_{\rm
  gas}$(S$=\!\!1$)) than suggested by thermal stability
arguments. \\*[-6cm] \includegraphics[width=9cm,
  angle=-90]{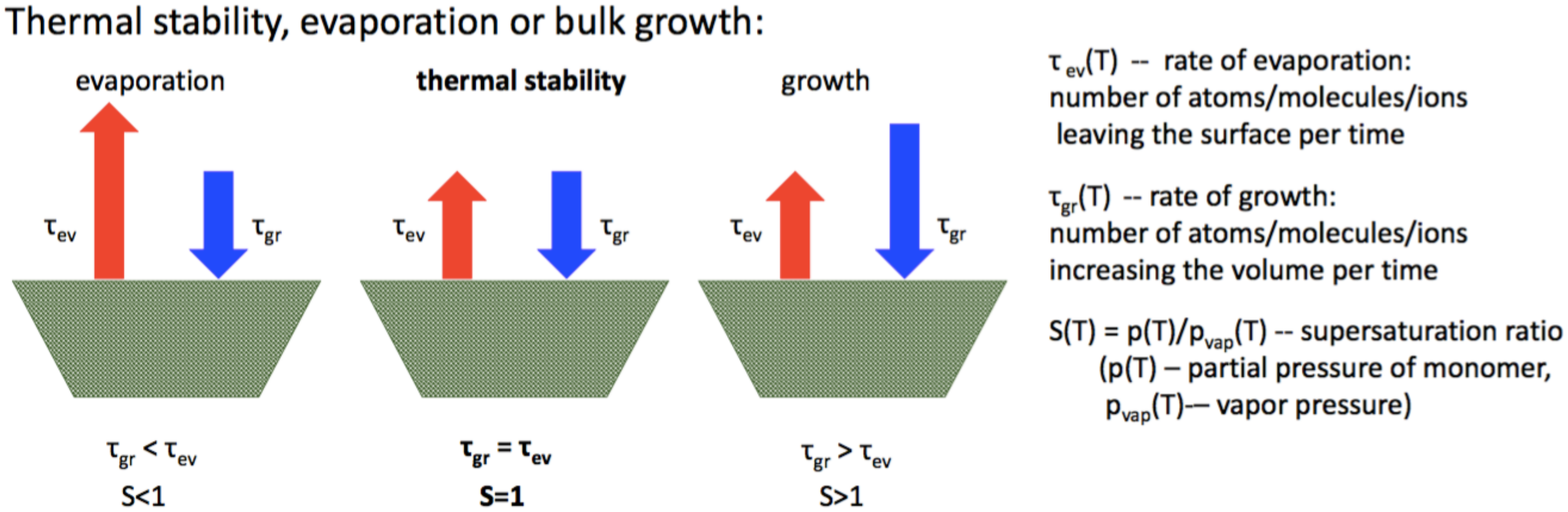}\\*[-6cm]
\end{textbox}

\smallskip
The formation of seed particles is a chain (or a network) of chemical
reactions that lead to an increasing complexity with each step: Larger
and larger molecules form that eventually are of the size of
clusters. Each of the reaction steps leads to the formation of the
next more complex cluster, e.g.  TiO$_2$ + TiO$_2$ $\longrightarrow$
(TiO$_2$)$_2$ + TiO$_2$ $\longrightarrow$ (TiO$_2$)$_3$ + TiO$_2$
$\longrightarrow$ $\ldots$ $\longrightarrow$ (TiO$_2$)$_{100}$
$\ldots$ $\longrightarrow$ TiO$_2$[s] (,[s]' stands for solid).  The
study of TiO$_2$-cluster formation suggests a chemical path where the
same molecule (monomer TiO$_2$) is added during each reaction step
(\citealt{2000JPhB...33.3417J}). Other species may not form that
easily from the gas phase as their monomers are not available as gas
species. Examples are silicates like Mg$_2$SiO$_4$, or metal oxides
like Al$_2$O$_3$.  (Al$_2$O$_3$)$_{\rm n}$ clusters have been observed
in extended gaseous envelopes of very cool, giants stars
(\cite{2017A&A...608A..55D} but the formation mechanisms is as of yet
uncertain. A further challenge is that monomers may exists in many
different geometrical configurations (isomers) that are equally
favorable (e.g. MgSiO$_3$, \citealt{2002CPL...363..145P}). Metal
clusters have also been studied but the example of iron is
discouraging as the (Fe)$_2$-cluster appears to be very unstable such
that the cluster growth chain of increasing complexity is interrupted
already at small cluster sizes.  The CLOUD (Cosmics Leaving Outdoor
Droplets) experiment at CERN (the European Organization for Nuclear
Research) has demonstrated that external radiation like through
high-energy particles from cosmic rays can support the formation of
condensation nuclei (\citealt{2016Sci...354.1119D}). Stellar radiation
(\citealt{2018arXiv180409054R}) and cosmic rays
(\citealt{2014IJAsB..13..173R}) ionise the high-altitude atmospheric
regions such that nucleation processes may involve complex ions in
exoplanets atmosphere at high altitudes.  But the locally very low gas
densities in the upper atmospheres may render the whole processes
inefficient for cloud or haze formation. The most
studied nucleation species is carbon which is only relevant for
exoplanets if the atmospheric gas is strongly oxygen depleted or
primordially carbon rich (\citealt{2014Life....4..142H}).

\subsubsection{Bulk growth/evaporation: mixed-material cloud particles with a non-homogeneous size distribution}\label{ss:bulkg}
The chemical diversity of exoplanet atmospheres allows for plenty of
materials being thermally stabile almost simultaneously.  Once
condensation seeds have been formed, these materials condense
simultaneously onto the seed's surface. Condensation of materials on
existing surfaces only requires a moderate supersaturation and is
therefore far easier occurring than the nucleation processes addressed
in Sect.~\ref{ss:nuc}.  Exoplanet atmosphere are expected to be
predominantly oxygen rich because all carbon is bound either in CO or
in CH$_4$ but CO$_2$ for hydrogen-poor gases. However, small rocky
planets may have lost their atmosphere or large fractions of their
atmospheric mass. Hydrogen, as the lightest element, would be the
first element to escape leaving behind a hydrogen-poor atmosphere with
an increased mean molecular weight ($\mu=\sum n_{\rm i}m_{\rm i}/\sum
n_{\rm i}$, $n_{\rm i}$ [cm$^{-3}$] - number density of gas species
$i$, $m_{\rm i}$ [g] -- mass of gas species $i$ ). Materials that form
the volume (and mass) bulk of the cloud particles in hydrogen-poor
(rocky) planets and hot, giant gas planets will therefore include
silicates (e.g. Mg$_2$SiO$_4$[s], MgSiO$_3$[s], SiO$_2$[s],
Fe$_2$SiO$_4$[s]), metal oxides (e.g. TiO$_2$[s], Al$_2$O$_3$[s],
MgO[s], FeO[s]) and others (e.g. Fe[s], CaTiO$_3$[s], FeS[s]).  If
Mg/Fe/Si/O are the most abundant species, as known to be the case for
the Sun, then these silicates will form the matrix of the cloud
particles that has mixed-in impurities from materials involving less
abundant elements (e.g. Ti and Al).  For colder exoplanets
(Earth-like, mini-Neptune), additional species may condense
(H$_2$O[l], NH$_3$[l], FeS[s]).  While the aforementioned solids
remain thermally stable also at low temperatures, and therefore keep
condensing, they may rearrange into more complex materials like
phyllosilicates as suggested by thermal equilibrium studies
(e.g. Mg$_2$SiO$_4$[s] $\longrightarrow$ Mg$_3$Si$_2$O$_9$H$_4$[s],
NaAlSi$_3$O$_8$[s] $\longrightarrow$
NaMg$_3$AlSi$_3$O$_{12}$H$_2$). The exact details are extremely
sensitive to the treatment of the element abundances in such
equilibrium considerations. A more robust result is that H$_2$O and
NH$_3$ condense at T$< 250$K, and that other species like NaCl[s],
FeS[s], Fe[s], Ni[s] will appear as impurities in such icy
cloud particles in exoplanet atmospheres. \cite{2017arXiv171201010W} have presented the most
stable condensation phases depending on local temperature and pressure
(phase diagrams) for all elements composing the solar element
abundance set including the effect of element depletion through the
condensation process. Such phase diagrams, however, do not allow to
determine the cloud particles sizes and a kinetic approach is required
for the actual calculations of the growth (and evaporation) rates for
each of the materials involved. Knowledge about the individual
reaction is required but laboratory data are only sparsely
available. For example, the formation of Mg$_2$SiO$_4$[s] alone can
involve 11 gas-surface reactions requiring Mg, Si, MgOH, Mg(OH)$_2$,
MgS, MgH, MgN, SiO, SiS, and H$_2$O from the gas phase. One efficient
method to include all these details in order to calculate cloud
particle properties (nucleation rates, material compositions, column
densities, particle sizes) is a moment method as described in
\cite{2013RSPTA.37110581H}. Another method involves the binning of the
particle size distribution
(\citealt{1997A&A...321..557K,2018arXiv180501468P}). Both methods were
also developed for dust formation modelling in AGB-stars envelopes.

\begin{figure}[h]
\vspace*{-3cm}
\hspace*{1.5cm}
\includegraphics[width=6cm, angle=-90]{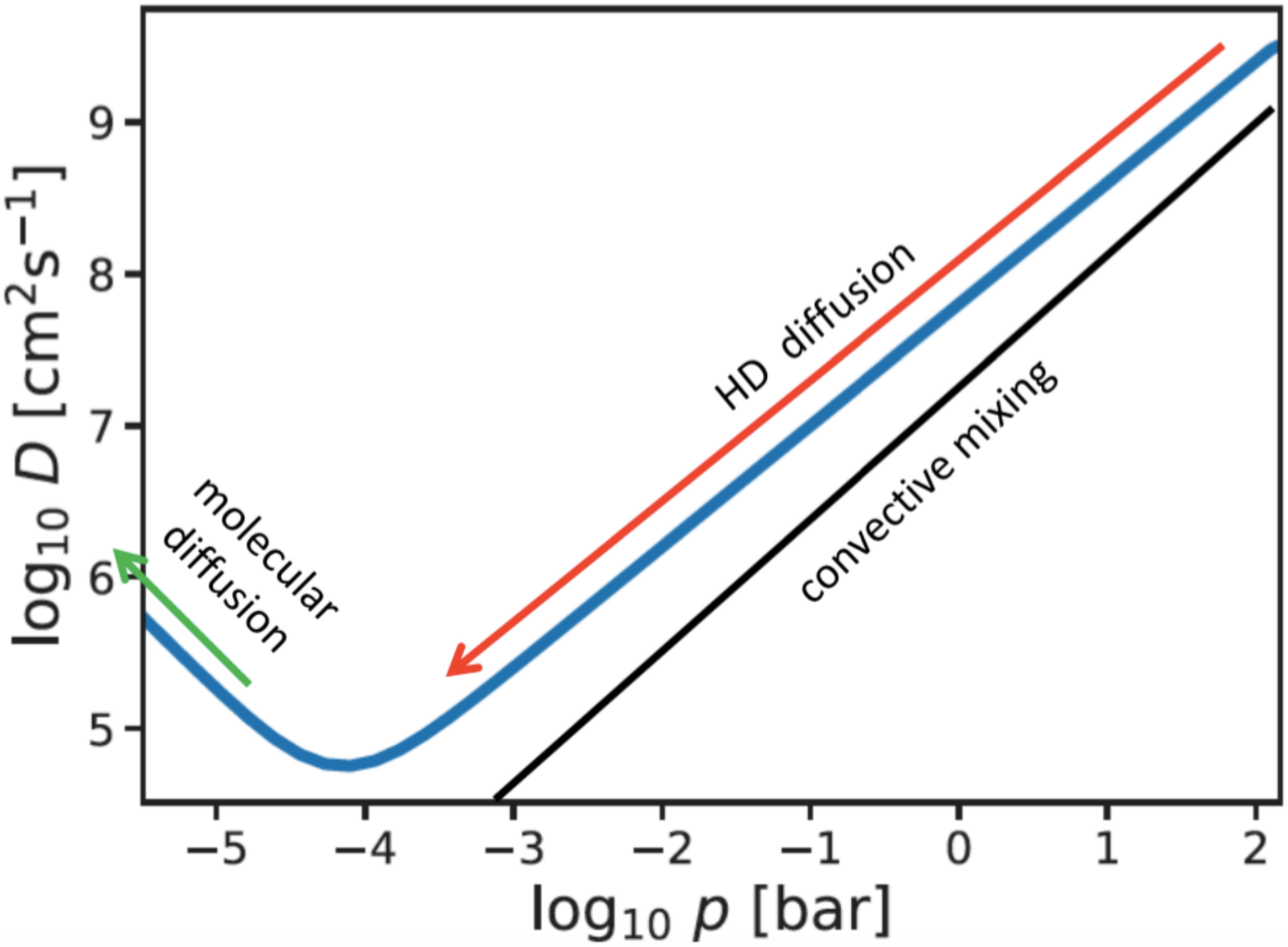}\\*[-3.2cm]
\caption{Element replenishment in exoplanet atmospheres can proceed through diffusive (blue line) and convective mixing (black line). Both decay with decreasing pressure, p [bar], but diffusion will pick up again  at the lowest pressures (green arrow) through molecular processes becoming more important than large-scale hydrodynamical processes (red arrow).  }
\label{fig_mix}
\end{figure}

\subsubsection{Gravitational settling and element replenishment}\label{ss:grav}

Cloud particles will move through the atmospheres with the gas if the
frictional coupling (drag) is strong, i.e., they would be swept along
with the wind. The cloud particles decouple from the gas if the drag
force is too small, hence, if the particles are too heavy
or the gas density is too low. This leads to the cloud particles
falling (gravitational settling) through the atmosphere with a speed
that is determined by their radius and material density, but also by
the surrounding gas density (\citealt{2003A&A...399..297W}). Rain
occurs if the cloud particles fall faster than the growth processes
can occur. In this case, their sizes remain constant until they
possibly evaporate. As the cloud particles move through the
atmosphere, they deplete the gas phase locally if they grow and they
enrich the local gas phase if they evaporate. Therefore, in order for
a largely continuous cloud coverage to be present in an atmosphere,
the gas needs to be replenished through hydrodynamical transport
processes. Such hydrodynamic motions include small-scale advection and
large scale convection, but gas replenishment can also proceed via the
much slower diffusion processes. The critical difference here is that
advection and convection transport a bulk of gas (fluid) with the
hydrodynamic (wind) velocity over a considerable distances
(e.g. day-night side equatorial jets,
\citealt{2008ApJ...682..559S,2013MNRAS.435.3159D,2014A&A...561A...1M}),
and that diffusion differentiates between the different gas species as
it depends on their masses (e.g. \citealt{2000Icar..143..244M}).
Figure~\ref{fig_mix} demonstrates that the diffusion reaches a minimum
when the large-scale motions become inefficient but picks up again at
lower densities due to the small-scale molecular diffusion
processes. A purely convective representation would drop in efficiency
with increasing distance from the convectively active regime.  The
bulk transport of gases as well as diffusive transport is part of the
solution of 3D hydrodynamic simulations. The challenge is to provide a
workable representation of element replenishment for 1D atmosphere
simulations which are far more efficient, in particular if cloud
formation, gas-phase chemistry and radiative transfer are a coupled
part of the complex model solution. So far, numerical experiments
testing diffusive material transport have been conducted for
cloud-free 3D simulations only (\citealt{2013A&A...558A..91P,
  2018arXiv180309149Z}). While being limited in their chemical
completeness, each numerical experiment is further limited by its
spatial (and time) resolution. This becomes critical for processes
predominantly acting on similar and smaller scales and it is known as
,turbulent closure problem'. Large efforts are going into representing
such sub-grid processes in Earth weather modelling
(e.g. \citealt{1990QJRMS.116..435S}) and in engineering
(e.g. \citealt{2015arXiv150604930G}). Turbulent, small-scale
fluctuations will extent the cloud formation regime for exoplanets and
amplifies the intermittency of the cloud coverage
(\citealt{2004A&A...423..657H}), but turbulence also facilitate cloud
particle charging in exoplanet atmospheres
(\citealt{2011ApJ...737...38H,2016PPCF...58g4003H}).


\begin{figure}[!htb]
\minipage{0.57\textwidth}
\vspace*{-0.7cm}
\hspace*{-1.5cm}\includegraphics[width=\linewidth]{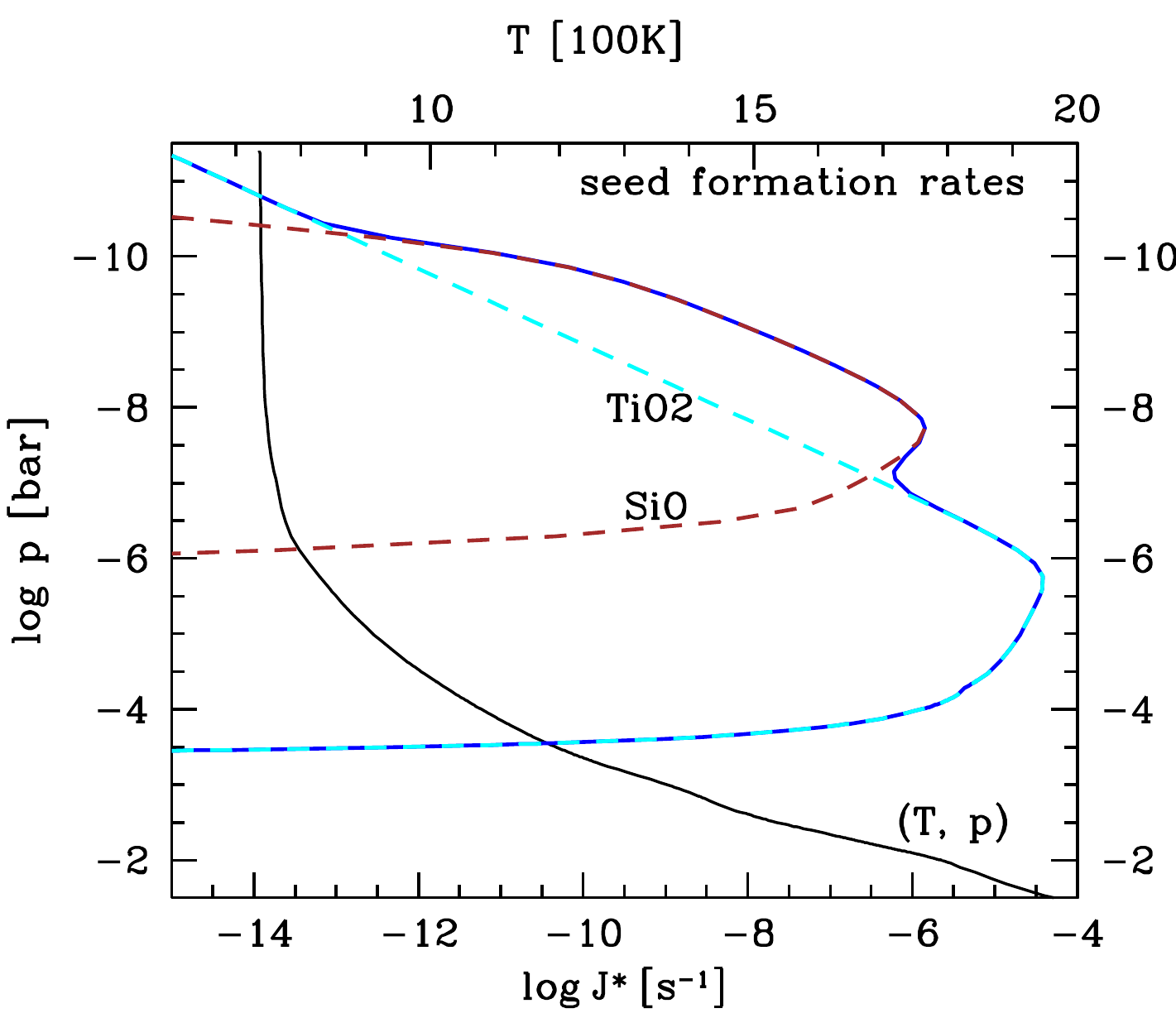}
\endminipage
\minipage{0.53\textwidth}
\hspace*{-1.5cm}\includegraphics[width=\linewidth]{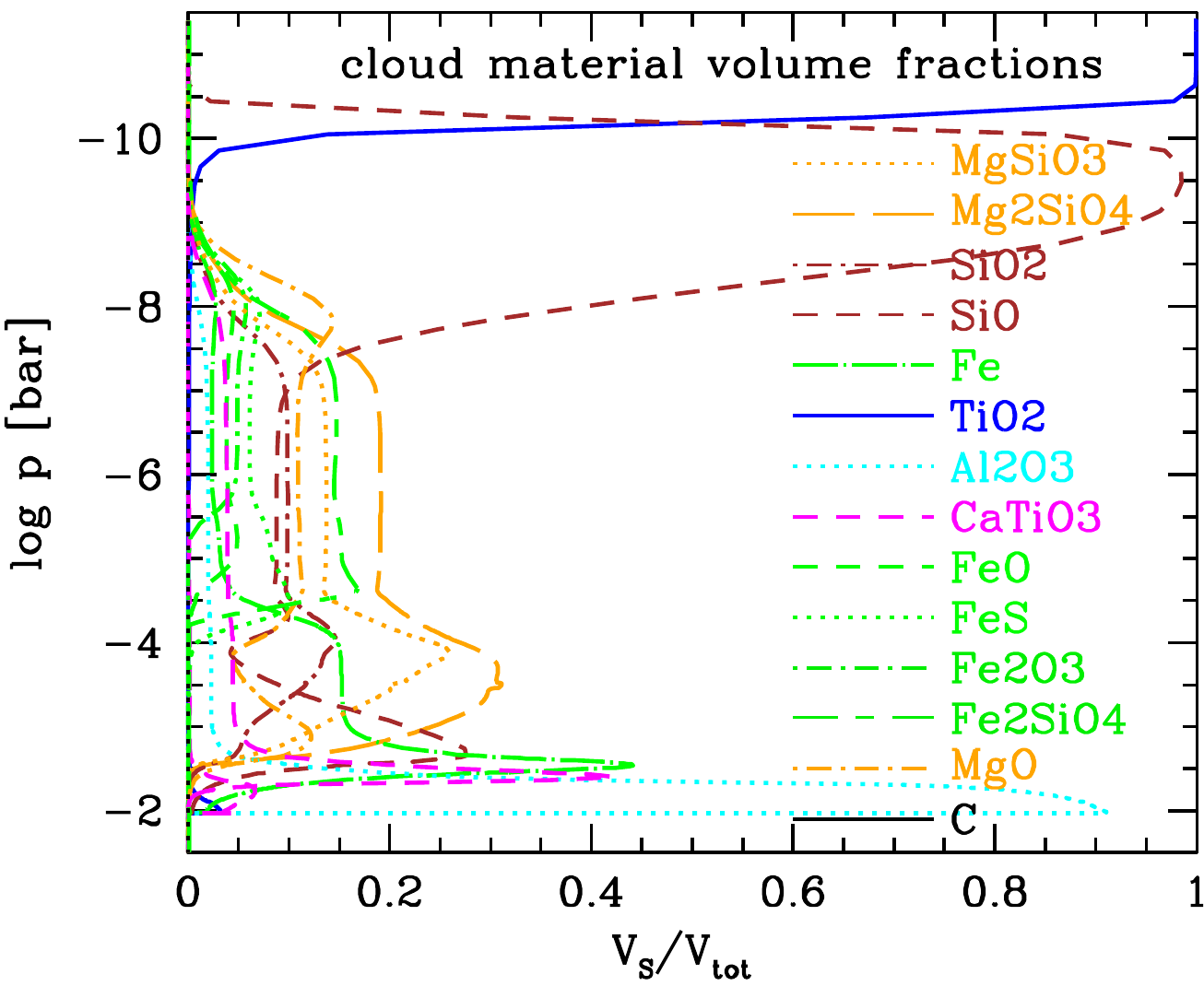}
\endminipage\\
\minipage{0.57\textwidth}
\hspace*{-1.5cm}\includegraphics[width=\linewidth]{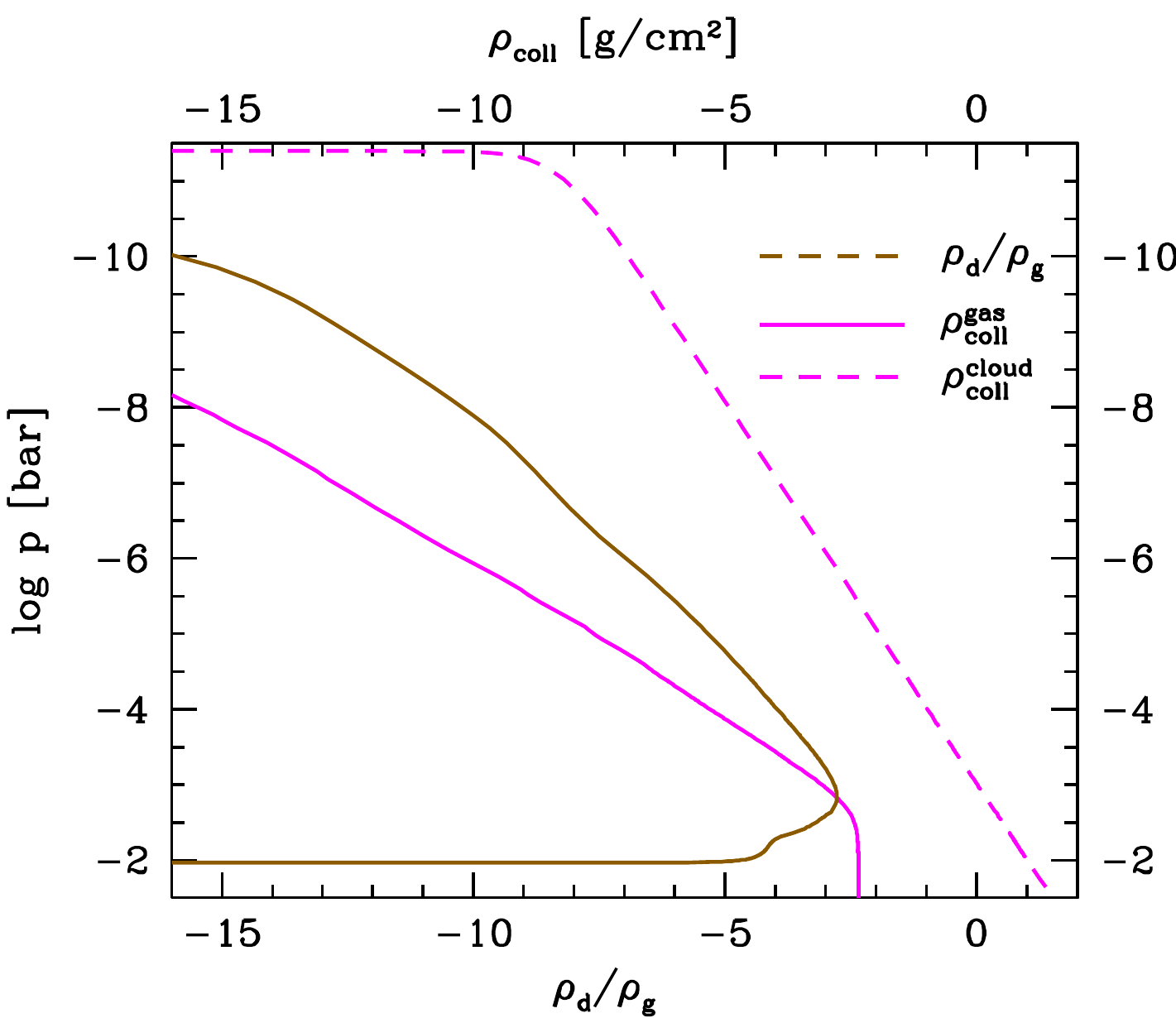}
\endminipage
\minipage{0.53\textwidth}
\vspace*{0.8cm}
\hspace*{-1.5cm}\includegraphics[width=\linewidth]{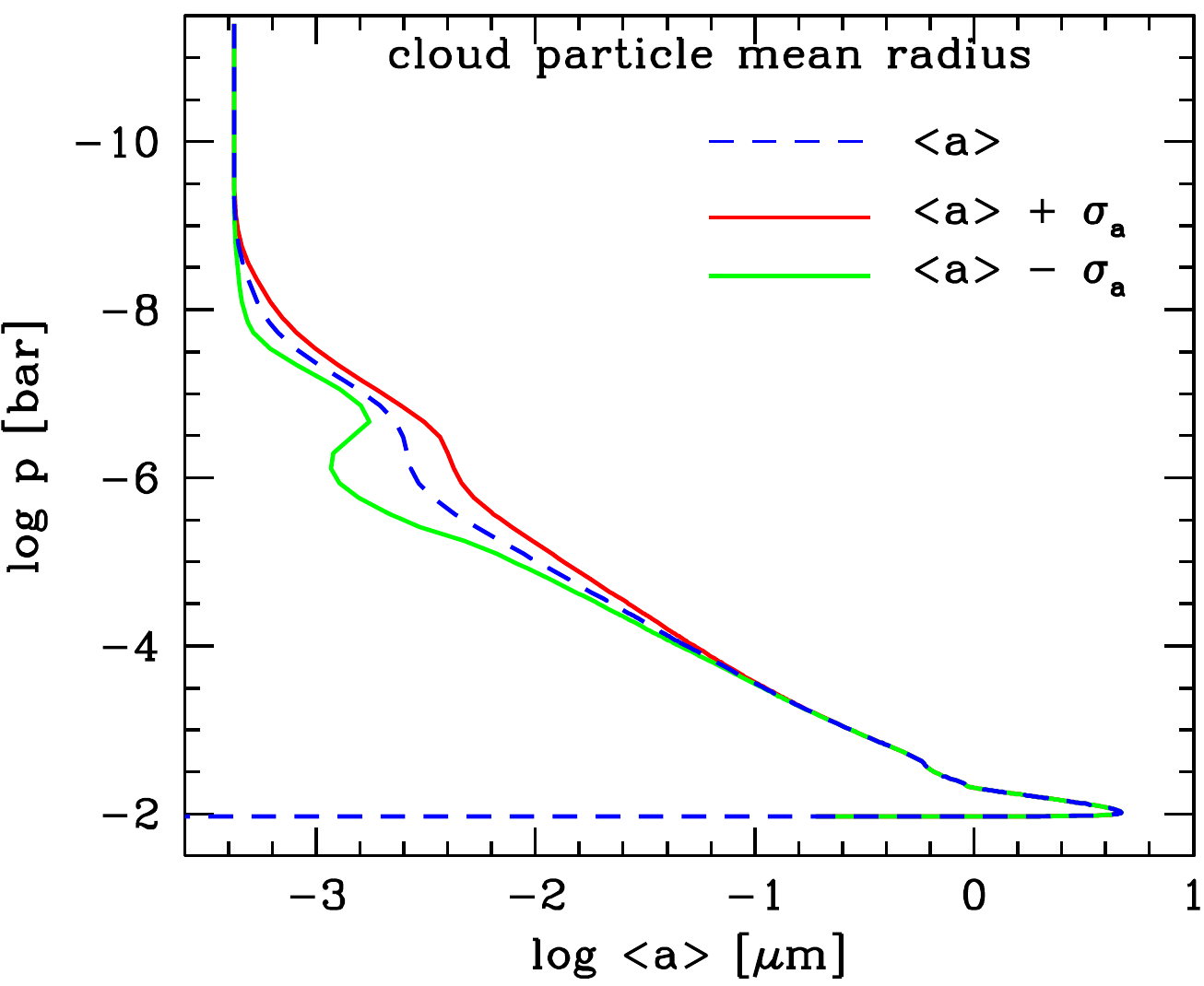}
\endminipage
\caption{Characteristic cloud properties in a hot exoplanet (T$_{\rm
    eff}$=1800K, log(g)=3.0, solar element abundances; (temperature T [K], pressure p [bar])-profile
  -- black solid line in top left). The nucleation rate (J$_*$ [s$^{-1}$], top left)
  determines the number of cloud particles unless transport processes
  sweep up condensation seeds from elsewhere. Exoplanet cloud
  particles are  made of a mix of materials (V$_{\rm s}$/V$_{\rm tot}$, s=TiO$_2$, Fe, MgSiO$_3$ etc, top right) which
  changes with atmospheric temperature (black line in top left), hence
  location. The dust-to-gas ratio ($\rho_{\rm d}/\rho_{\rm gas}$)
  demonstrates where the main cloud mass is located. It is
  correlated to the cloud column density ($\rho_{\rm coll}^{\rm
    cloud}$; both bottom left). The mean cloud particle sizes ($\langle a \rangle$ [$\mu$m]) are
  different at different atmospheric heights (dashed blue line in
  bottom right) ranging from $10^{-6}$cm at the top of the cloud to  $>0.5\mu$m at the cloud bottom.  Local particle sizes deviate from this mean value ($\sigma_{\rm a}$ -- standard deviation; red and green lines) most in the area of strongest surface growth $\sim 10^{-6}$ bar. The largest particles reside in the
  densest part of the atmosphere and are made of high-temperature
  condensates like CaTiO$_3$[s] and Al$_2$O$_3$[s]. }
\label{fig3}
\end{figure}

\section{A generic case study for exoplanet clouds in 1D}
Cloud formation is determined by the local thermodynamic state (gas
temperature and gas pressure) of the atmospheric gas. The
thermodynamic state will be affected by the planet size and the
external radiation from the host star (lava planets vs. giant gas
planets) or the interplanetary medium ('directly imaged planets',
i.e. planets with a large orbital distance from their host star), by
the interaction between the atmosphere and the crust (rocky planets),
and by the evolutionary state (young vs old) of the planet
itself. Because of the large diversity of exoplanets, it is desirable
to develop a fundamental understanding about their atmospheres and
about exoplanet clouds. We therefore choose a generic case to discuss
predictive results. Our example represents a moderately warm giant gas
planet (directly imaged and/or self-luminous young planet) with
effective temperature T$_{\rm eff}$=1800K, surface gravity log(g)=3.0,
an initial set of solar element abundances (i.e. the number of
elements of one kind contained in all gas species, e.g. Si in SiO,
SiH, SiO$_2$, MgSiO$_3$, SiS etc., or Fe in Fe, FeO, FeS, FeSiO$_3$
etc.) and with enough distance from its host star that stellar
irradiation does not matter.  The results are obtained from
considering the formation of three nucleation species simultaneously
(TiO$_2$, SiO, C), and the formation of 15 bulk materials (listed in
Fig.\ref{fig3}) that form from 9 elements (Mg, Si, Ti, O, Fe, Al, Ca,
S, C) by 126 surface reactions. The approach is based on
\cite{2008A&A...485..547H} and follows the principles described in
Sect.~\ref{s:how}.  Figure~\ref{fig3} presents the exoplanet cloud
results for a set of characteristic properties :\\ -- the seed
formation rate, J$_*$ [s$^{-1}$] (top left panel),\\ -- the material
volume fractions, V$_{\rm s}$/V$_{\rm tot}$ (V$_{\rm s}$ -- local
volume fraction of the material s,\\ V$_{\rm tot}$  -- total, local cloud
volume, s - solid material like TiO$_2$[s], SiO[s] etc.; top right
panel), \\ -- the cloud and gas column densities, $\rho_{\rm col}$ [g
  cm$^{-2}$] , the cloud mass load of the atmosphere in terms of the
ratio between cloud mass density and gas mass density, $\rho_{\rm
  d}/\rho_{\rm g}$, and the cloud column density, $\rho_{\rm
  coll}^{\rm cloud}$ (bottom left),\\ -- and the mean cloud particle
radii, $\langle a\rangle$ [$\mu$m] (bottom right). \\ The cloud column
density ($\rho_{\rm col}$ [g cm$^{-2}$]) is the integrated local cloud
particle number density (number of cloud particles per volume) over
the geometrical extension of the cloud.  The number of particles that
make up the whole cloud is determined by the seed formation rate
(J$_*$ [s$^{-1}$]). Both, TiO$_2$ and SiO form condensation
seeds. Carbon does not form as we consider a gas that has more oxygen
than carbon (C/O=0.53). SiO seed formation is most efficient at lower
temperature/pressures and TiO$_2$ picks up at higher temperatures.
The whole region of efficient seed formation (J$_*>10^{-8}$ s$^{-1}$)
spans about 4 orders of magnitude in pressure in the upper atmospheric
regions with $p<10^{-4}$bar. (Note that the same panel shows the
atmospheric (T, p) structure in black on the right axis for
comparison.). The emergence of the condensation seeds is also apparent
from the panel showing the material composition of the cloud particles
(top right): The uppermost, low pressure regions are made solely of
very small TiO$_2$[s] and SiO[s] condensation seeds ({\it haze}). The
bulk of the cloud particle volume forms through condensation (surface
reactions) of all materials at $p>10^{-8}$bar. At the cloud centre
($p\approx10^{-6}$bar) cloud particles are made of a mix of
Mg$_2$SiO$_4$[s] (20\%), Fe$_2$SiO$_4$[s] (17.5\%) and MgSiO$_3$[s]
(16.5\%), and MgO[s], SiO$_2$[s] and SiO[s] form circa 15\% each of
the total cloud material. All other materials occur as inclusions with
$\lesssim$ 10\%. Silicate materials evaporate at $p\approx10^{-3}$bar
(here: T$\approx$ 1700K). The innermost cloud layers are made of
high-temperature condensates (materials that are thermally stable at
high temperatures) like Fe[s], CaTiO$_3$[s] and Al$_2$O$_3$[s].  The
consistent solution of the cloud formation processes allows us now to
link the cloud material composition with the resulting cloud particle
sizes (lower right). We start our discussion again in the low pressure
upper atmosphere region which is dominated by the nucleation
process. Here we find that the cloud particles remain very small
($\approx 10^{-3}\mu$m) until exactly where the bulk growth sets in
efficiently. The mean particles size, $\langle a\rangle$ (dashed blue
line), increases steadily inwards because gravitational settling
causes the cloud particles to fall into atmospheric regions of higher
densities but also of increasing temperature.  Note that the cloud
expands beneath the nucleation zone. The high densities accelerate the
surface growth processes leading to the steep inward increase of the
cloud particle's mean radius (and their fall speed; not shown) until
they encounter the temperatures that cause complete evaporation. The
inwards increasing particle sizes do cause an increase of the
atmosphere's cloud mass load (lower left, brown dashed line): The
ratio between the cloud mass density and the gas density, $\rho_{\rm
  d}/\rho_{\rm g}$ (so-called 'dust-to-gas ratio'), is largest where
the mean particle sizes reach their maximum.

Cloud particles develop not only a vertical size distribution f(a(z))
(blue dashed line, lower right panel Fig.~\ref{fig3}), but appear
with a local size distribution f(a(z), z). This can, for example, be
represented by a Gaussian distribution with a standard deviation,
$\sigma_{\rm a}(z)$ (solid green and red line in lower right
panel)\footnote{For this paper, the height dependent parameters,
  $N(z,a)$ and $\sigma(z)$, for a Gaussian distribution, $f(a,
  N)=N/(2\pi^2\sigma)\cdot \exp(- [ (a - \langle
    a\rangle)/\sigma]^2)$, are derived using dust moments $L_{\rm
    j}(z)=\int_{\rm V_l}^{\infty} V(z)^{j/3}f(V,z)dV$ as described in
  \cite{2008A&A...485..547H}. $N(z,a)$ is the number of cloud
  particles at a certain height z=z(p,T) of a certain size $a$, and
  $\sigma(z)$ is the local standard deviation, $V(z)$ is the local
  cloud particle volume.}. The local size distributions are very
narrow in the low-density cloud regions ($p<10^{-9}$bar for the model
shown) and at the cloud base ($p>10^{-4}$bar for the model
shown). The size distribution broadens when the bulk growth sets
in. The width of the size distribution is also determined by the
nucleation process. It is the widest when the second nucleation peak appears due
to the inset of efficient  TiO$_2$ nucleation  ($p\approx10^{-6}$bar for the model shown). At this
location, the size distribution contains a considerable number of
smaller cloud particles. In conclusion, in order to present the local
cloud particle size distribution, all size determining processes need
to be consistently treated. \cite{2018arXiv180501468P} do present
results for double-peaked local particle size distribution based on a
binning method.

\begin{figure}[h]
\hspace*{-1cm}
\includegraphics[width=10cm]{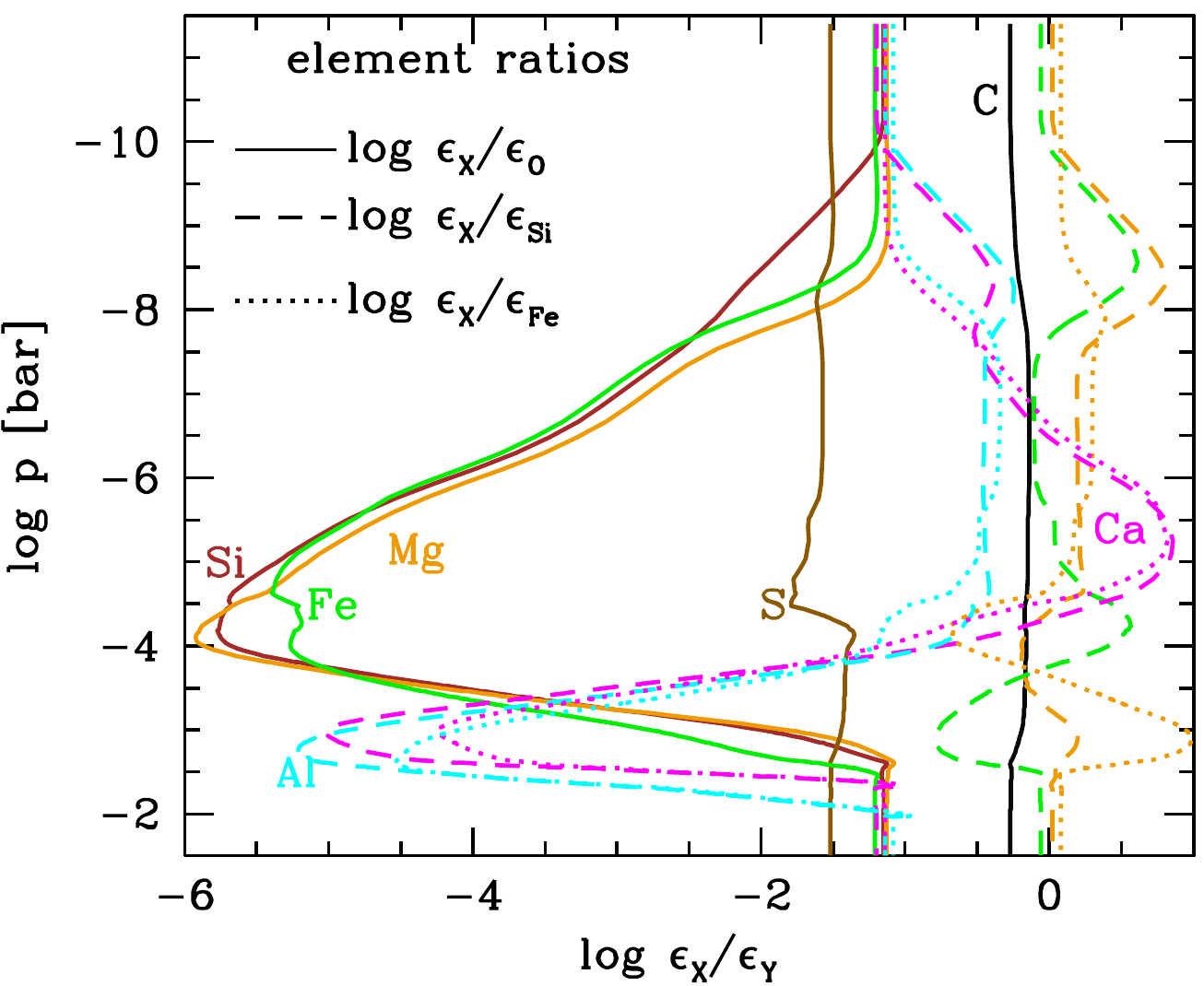}
\caption{Element abundance ratios (mineralogic ratios) in relation to oxygen (O, solid
  lines), silicon (Si, dashed lines) and iron (Fe, dotted lines) on a
  logarithmic scale for the same model like in Fig.~\ref{fig3}. The
  individual elements are colour coded. The C/O ratio is indicative of
  an oxygen rich gas with the unperturbed, solar value of C/O=0.537 above and below the cloud. The maximum value reached in
  this giant gas planet atmosphere model is C/O=0.725. }
\label{figElm}
\end{figure}

\subsection{Thoughts on model completeness} The question arises in how far
the above drawn picture of clouds on exoplanets is complete.  It is
incomplete in the sense that not all possible thermally stable
materials have been included and that only those resulting from the
most abundant (solar) elements were considered. The alternative to
this is to apply a phase equilibrium (thermal equilibrium; see
\citealt{2015ARA&A..53..279M}) approach which, however, will not
provide information about, for example, particle sizes nor knowledge
about if these materials would actually form for a given atmospheric
condition. We have derived the principle characteristics for exoplanet
clouds based on a 1D atmosphere configuration. This naturally neglects
all multi-dimensional effects like equatorial jets, turbulence, or
scattering of stellar photons beyond the day-night terminator, but it
is the only way to build our understanding based on a consistent
theoretical approach.  This knowledge can then be applied to 3D
radiation-hydrodynamics models including all the required gas and
cloud chemistry (\citealt{2016A&A...594A..48L,2018arXiv180300226L}),
or inform retrieval models. Retrieval models necessarily use a 1D
atmospheric profile that require heavy parameterisation to allow the
simulation of grids of $10^6$ atmosphere structures. However, as it
turns out, the parameterised temperature profiles resemble 3D
structures well if enough parameters are allowed and if high
resolution and more data points are included
(\citealt{2017ApJ...848..127B}).  Furthermore, exoplanet atmospheres
are exposed to external radiation from the host star but also from
high-energy cosmic rays leading to atmospheric electrification
(\citealt{2016SGeo...37..705H}). The host-star's UV field ionises the
upper atmospheric regions for close-in planets and so does the
interplanetary radiation field for the more distant planets in a
planetary system. Cosmic ray particles arrive with a far higher
energy, though on a lower rate, and reach deeper atmospheric layers or
even the planetary surface where they trigger the appearance of HCN
(\citealt{2016NatGe...9..452A}). In addition, photon-initiated chemistry (oppose to chemical reactions occurring due to the collision between two gas species) occurs in
the outermost layers of an exoplanet atmosphere. The formation of
photochemical haze is form of large carbo-hydrate molecules is discussed to require a  combination of gas-phase photochemistry and coagulation (collisional reaction between exiting particles) of photochemical educts  which then produce
so-called tholins (\citealt{2018ApJ...853....7K}). Tholins (\citealt{1979Natur.277..102S}), describe a variety of organic compounds that are not specified any further.

\subsection{Summary of take-away points about clouds in exoplanet atmospheres}
\begin{enumerate}
\item Small, haze-sized aerosols constitute the upper boundary of clouds. 
\item The seed formation rate determines the number of cloud particles.
\item Any middle part of a cloud will be chemically very rich. Cloud  particles are made of a mix of all thermally stable materials. An optically thick atmosphere results.
\item  Cloud particle radii and their material composition change throughout the cloud because of changing thermodynamic conditions.
\item Large cloud particles made of high-temperature condensates form the cloud base.
\end{enumerate}

\begin{figure}[!htb]
\vspace*{-2cm}
\hspace*{-0cm}\includegraphics[width=13cm]{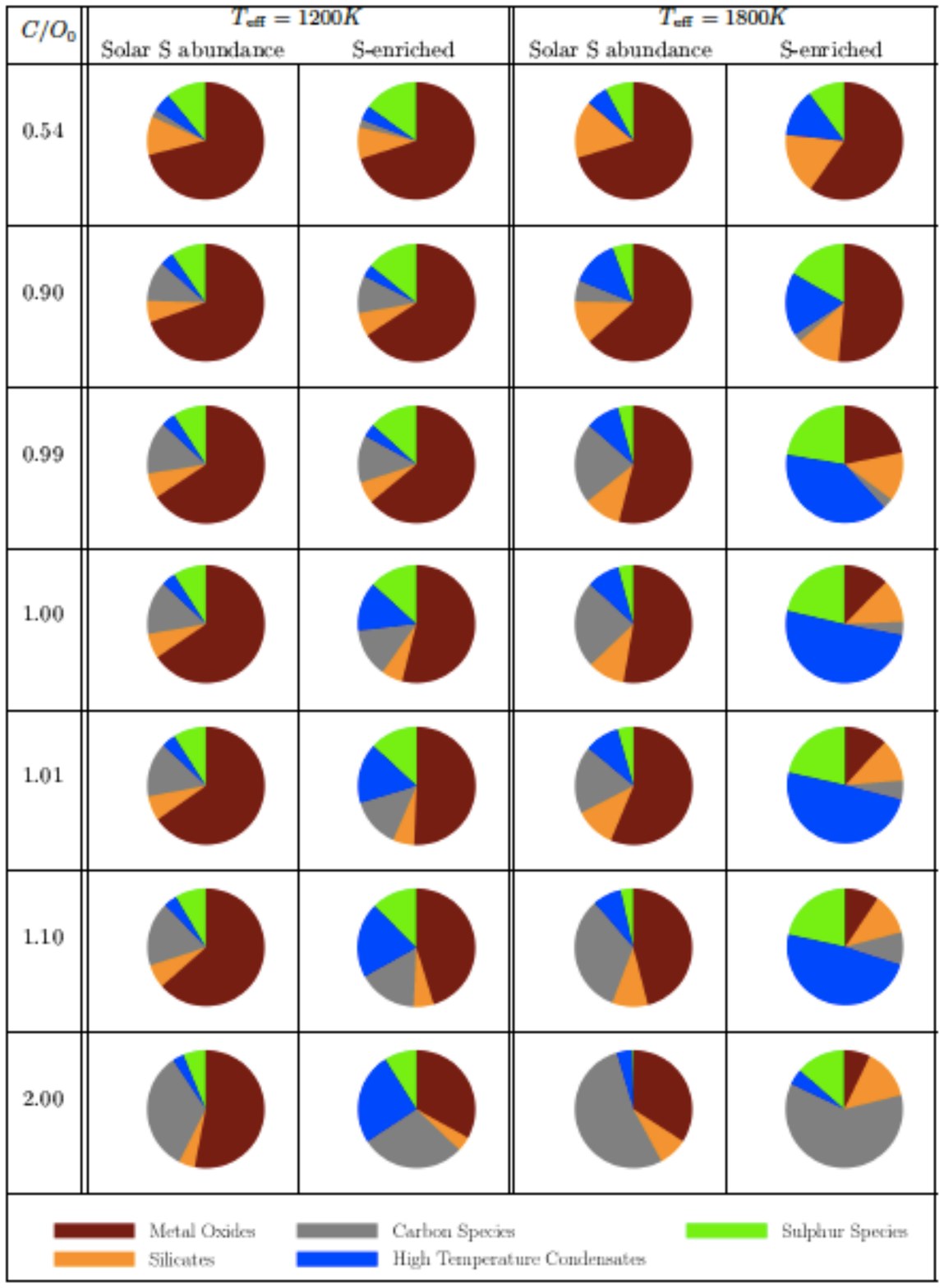}\\*[-2.5cm]
\caption{The cumulative cloud material volume fractions for a set of
  1D {\sc DriftPhoenix} model atmospheres for varying carbon-to-oxygen ratio (C/O) and
  sulphur abundance. The focus of this figure is the transition from
  an oxygen-rich to a potentially carbon-rich atmosphere: Carbon
  condensation occurs already at C/O=0.9 (and lower for lower T$_{\rm
    eff}$) but is suppressed upto C/O$\approx$1.1 by an increase of sulphur in particular
  in warmer atmospheres.  The cloud material
  species were combined in 5 groups: metal oxides (SiO[s], SiO$_2$[s],
  MgO[s], FeO[s]), silicates (MgSiO$_3$[s], Mg2SiO$_4$[s]), carbon
  species (C[s], SiC[s], TiC[s]), high temperature condensates
  (TiO$_2$[s], Al$_2$O$_3$[s], CaTiO$_3$[s], Fe[s]), and sulfur
  species (S[s], MgS[s], FeS[s]).}
  \label{figpies}
\end{figure}

\section{Exoplanet element abundances and mineralogic ratios}

The original element abundances (i.e. the number of elements of one
kind contained in all gas species which has not yet been affected by
cloud formation) that determine the chemical composition of an
exoplanet's atmosphere gas may reflect its formation processes
(e.g. \citealt{2014Life....4..142H,2015ApJ...813...72C,2016MNRAS.461.3274C,2016A&A...595A..83E,2017MNRAS.469.4102M}). Hence,
attempts are ongoing to use element abundance measurements from
exoplanet atmospheres to constrain planet formation processes
(e.g. \citealt{2016MNRAS.463.4516P}). It will, however, be challenging
to determine the values of the initial element abundances from
spectroscopy of the exoplanet atmosphere if clouds have formed.  Cloud
nucleation and growth consume elements, therefore causing specific
atoms and molecules to drop in abundance. Transport processes like
gravitational settling or advection will de-localise the process.
Linking the observed atmospheric exoplanet element abundances to the
host star's element abundances is therefore a non-trivial task, unless
a complex (i.e. as complete as possible) modelling approach is
adopted. It appears more feasible to address the relation between the
host star's element abundances and the element abundance of the
planet's core (e.g. \citealt{2017A&A...608A..94S}).  The challenges
hereby, however, lay with the precise measurement of the stellar
properties as demonstrated for example for 55 Cnc e
(\citealt{2018arXiv180407537C}). Matters are further complicated if
the element abundances are derived from observations of molecules that
are (additionally) affected by photochemistry (e.g. CO/CH$_4$,
N$_2$/NH$_3$) in the atmospheric gas above the cloud layer
(e.g. \citealt{2016ApJ...829...66M}) or cosmic ray impact
(\citealt{2014IJAsB..13..173R}).  Figure~\ref{figElm} summarises the
element depletion by cloud formation in terms of element ratios (also
called mineralogic ratios), $\log \epsilon_{\rm X}/\epsilon_{\rm O}$
(solid lines), $\log \epsilon_{\rm X}/\epsilon_{\rm Si}$ (dashed
lines), and $\log \epsilon_{\rm X}/\epsilon_{\rm Fe}$ (dotted lines)
for the same exoplanet atmosphere as in Fig.~\ref{fig3}. The most
interesting is the carbon-to-oxygen ration (C/O) which varies between
the initial solar value of $\approx 0.54$ and $\approx 0.73$. This
emphasises that there is no one C/O that can be used to
characterise a cloud-forming exoplanet. Such a change of C/O due to
cloud formation can convert the gas-phase chemistry from an
oxygen-dominated into a carbon-dominated gas if C/O  is
already $\geq0.85$ as result of planet formation processes
(\citealt{2014Life....4..142H,2017MNRAS.469.4102M}). Molecules like
e.g. C$_2$H$_2$, HCN, CH and polycyclic-aromatic-hydrocarbon (PAH) molecules can then be expected to
appear in the exoplanet absorption spectrum
({\citealt{2013MNRAS.435.1888B}) where photochemistry is
  inefficient. The ratios Fe/O, Si/O and Mg/O are rather similar as
  the depleting materials are thermally stable at similar
  temperatures, and Fe/Si/Mg have comparable initial element abundances. Sulphur appears to play no substantial role for
  element depletion in warm giant gas planets. The Mg/Si maximum is
  $\approx 1.56$, the minimum is $\approx0.64$ compared to a solar
  value of 1.05. The Fe/Si maximum is $\approx 3.66$, its minimum
  $\approx0.16$ compared to a solar value of 0.87. We note that the
  precise local values of the mineralogic values will depend on
  whether the element replenishment is convective or diffusive. The
  maximum/minimum values cited here should, however, be independent of
  the element replenishment mechanism. 

\begin{figure}
\minipage{0.57\textwidth}
\vspace*{-0.7cm}
\hspace*{-1.5cm}\includegraphics[width=\linewidth]{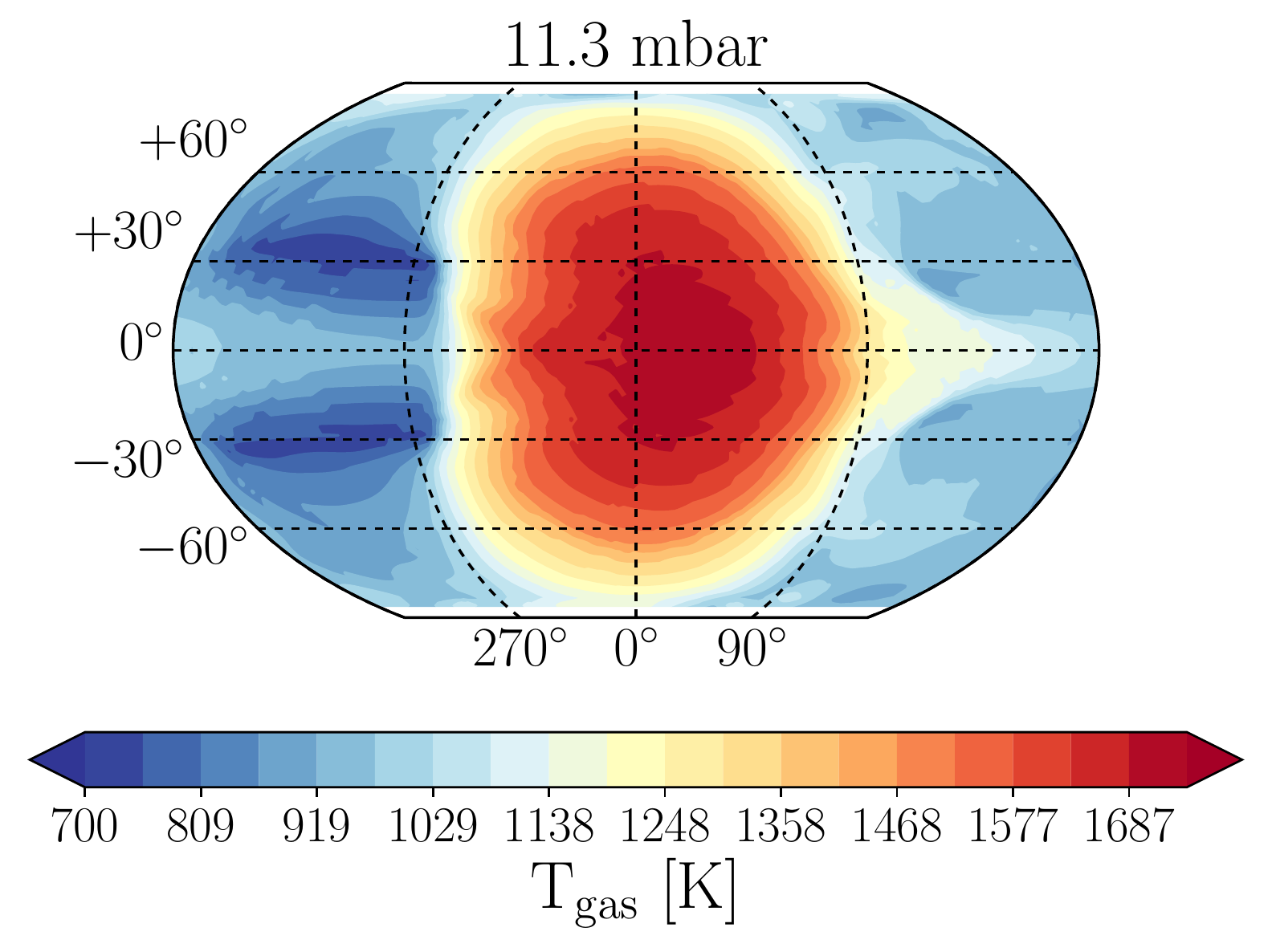}
\endminipage
\minipage{0.58\textwidth}
\hspace*{-1.7cm}\includegraphics[width=\linewidth]{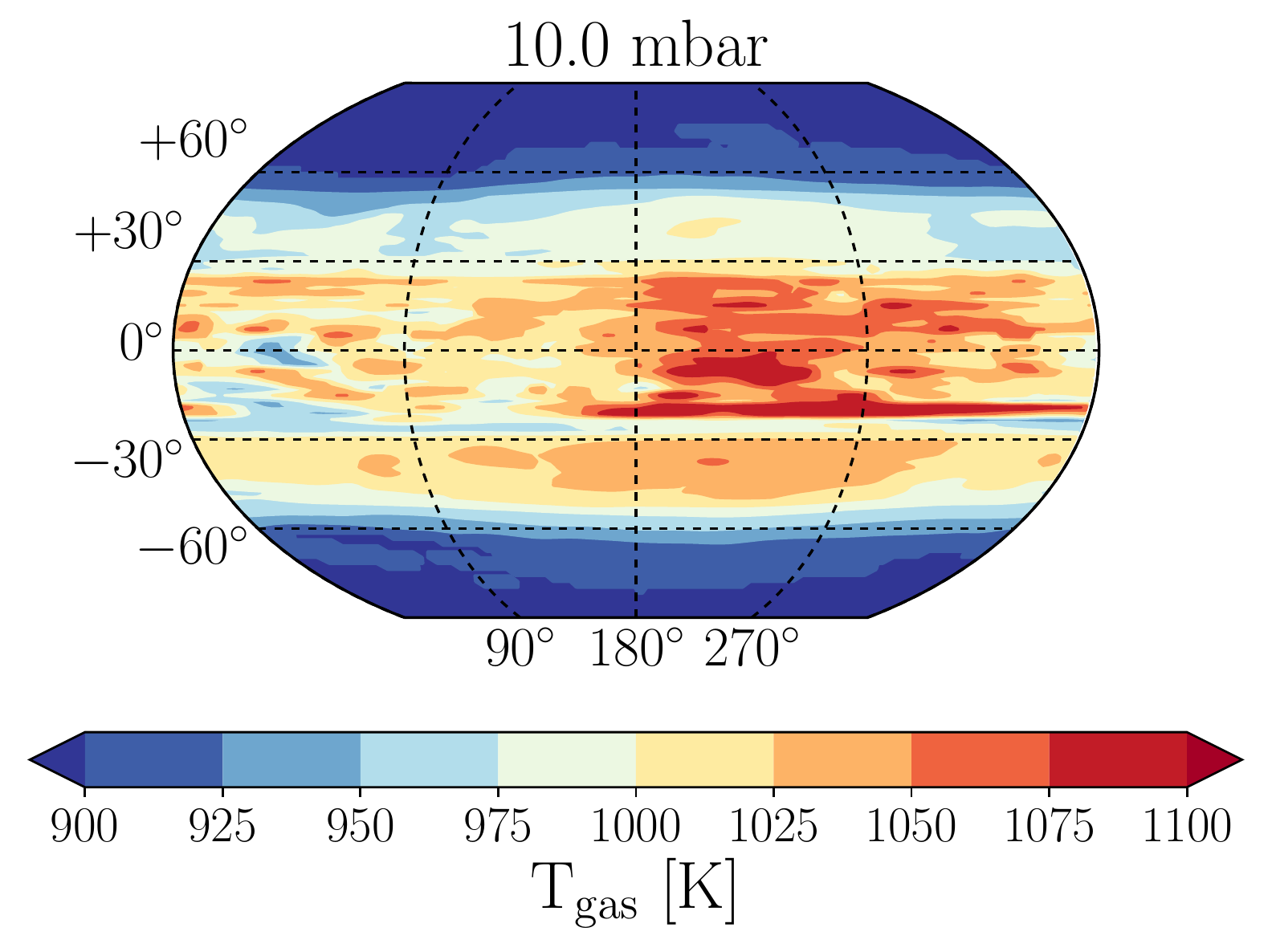}
\endminipage\\
\minipage{0.57\textwidth}
\hspace*{-1.7cm}\includegraphics[width=\linewidth]{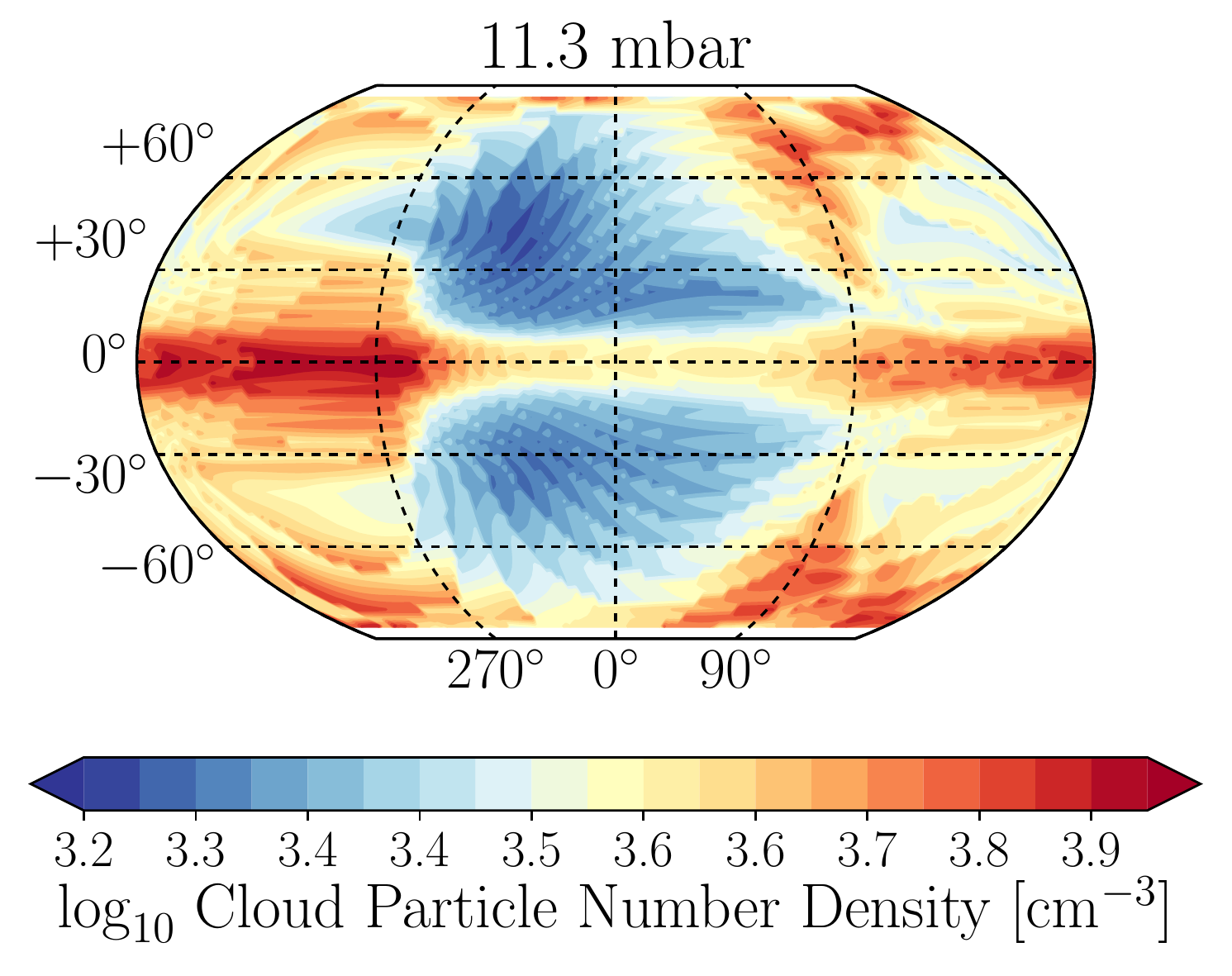}
\endminipage
\minipage{0.58\textwidth}
\hspace*{-1.7cm}\includegraphics[width=\linewidth]{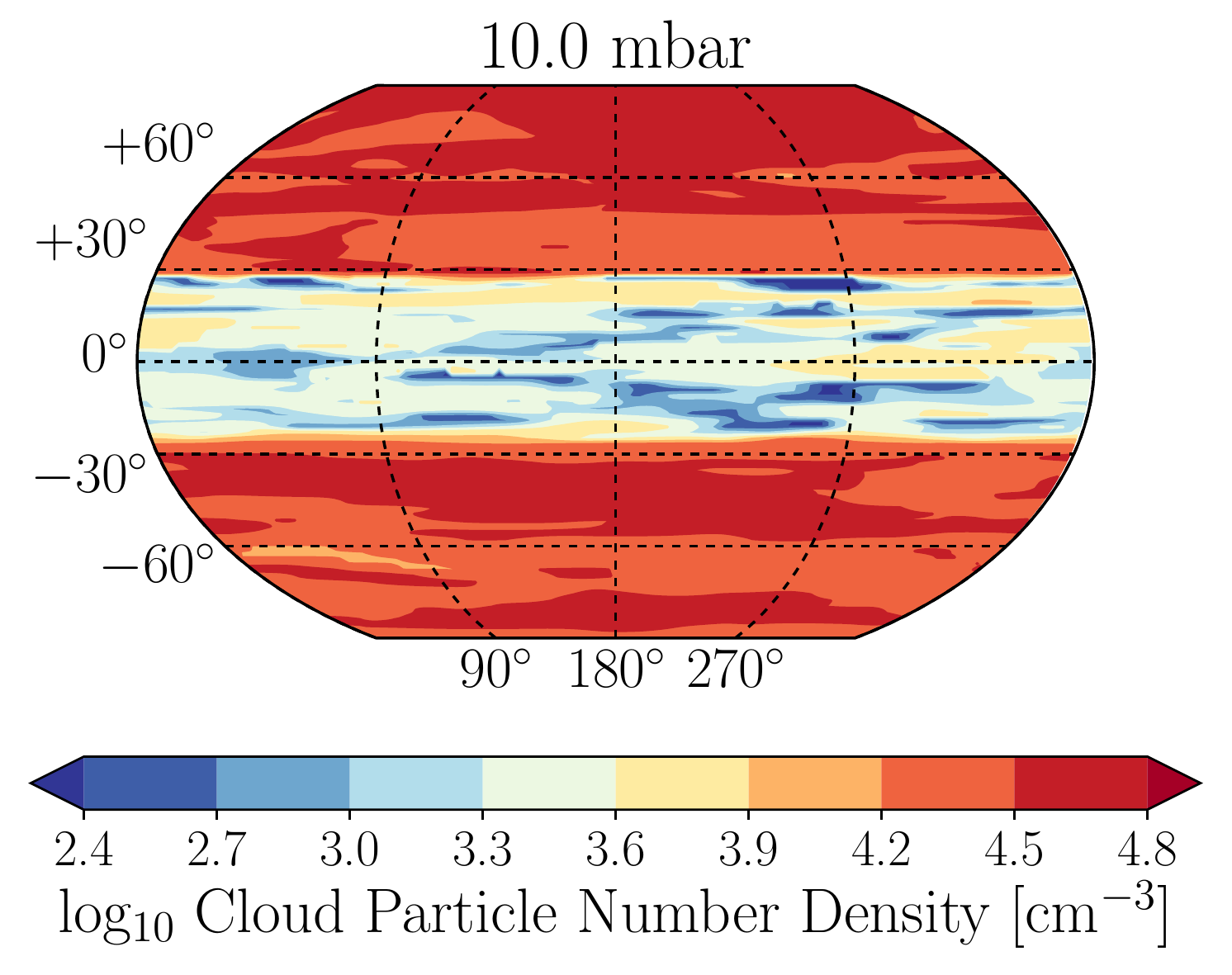}
\endminipage\\
\minipage{0.57\textwidth}
\hspace*{-1.5cm}\includegraphics[width=\linewidth]{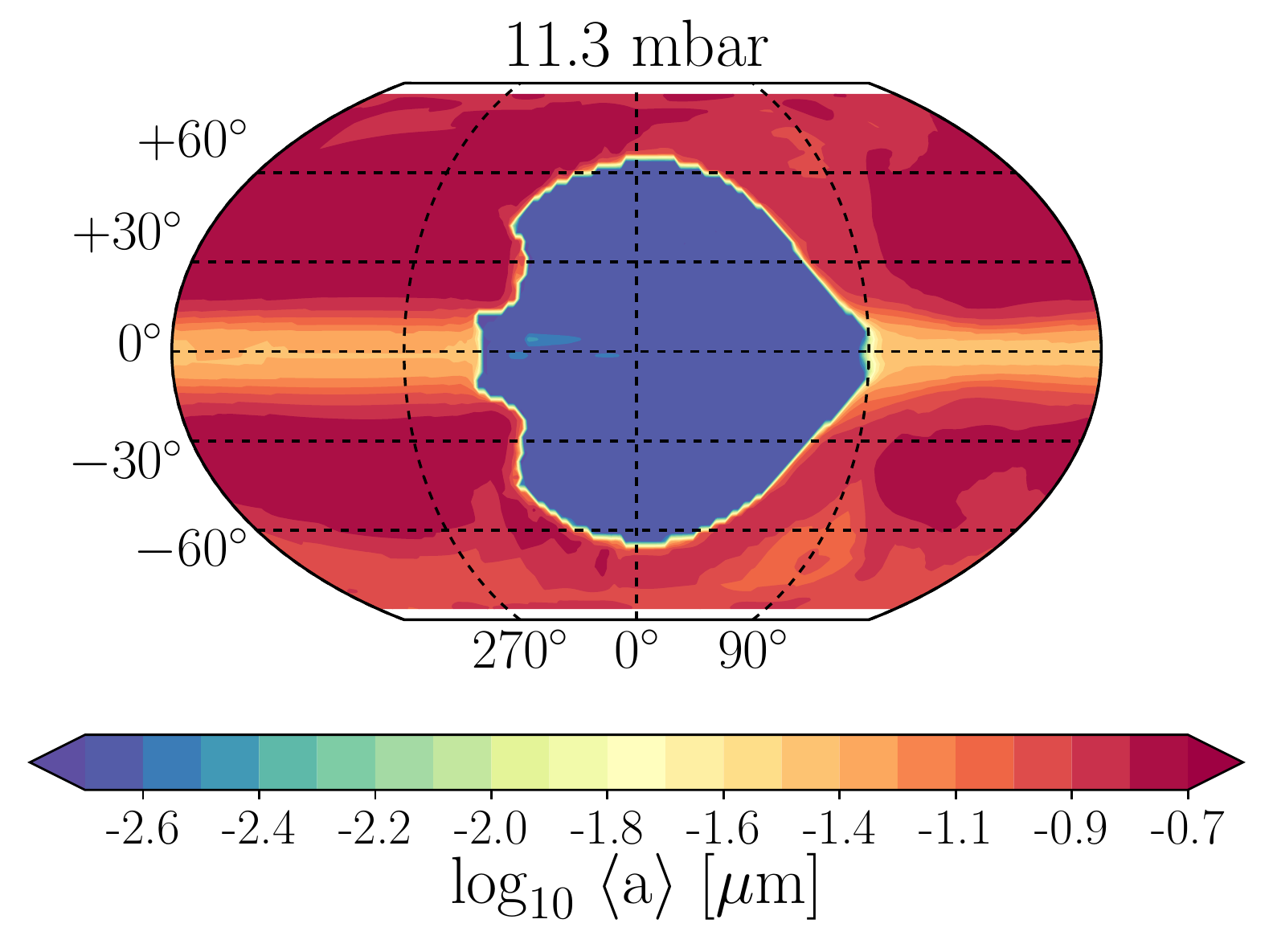}
\endminipage
\minipage{0.58\textwidth}
\hspace*{-1.7cm}\includegraphics[width=\linewidth]{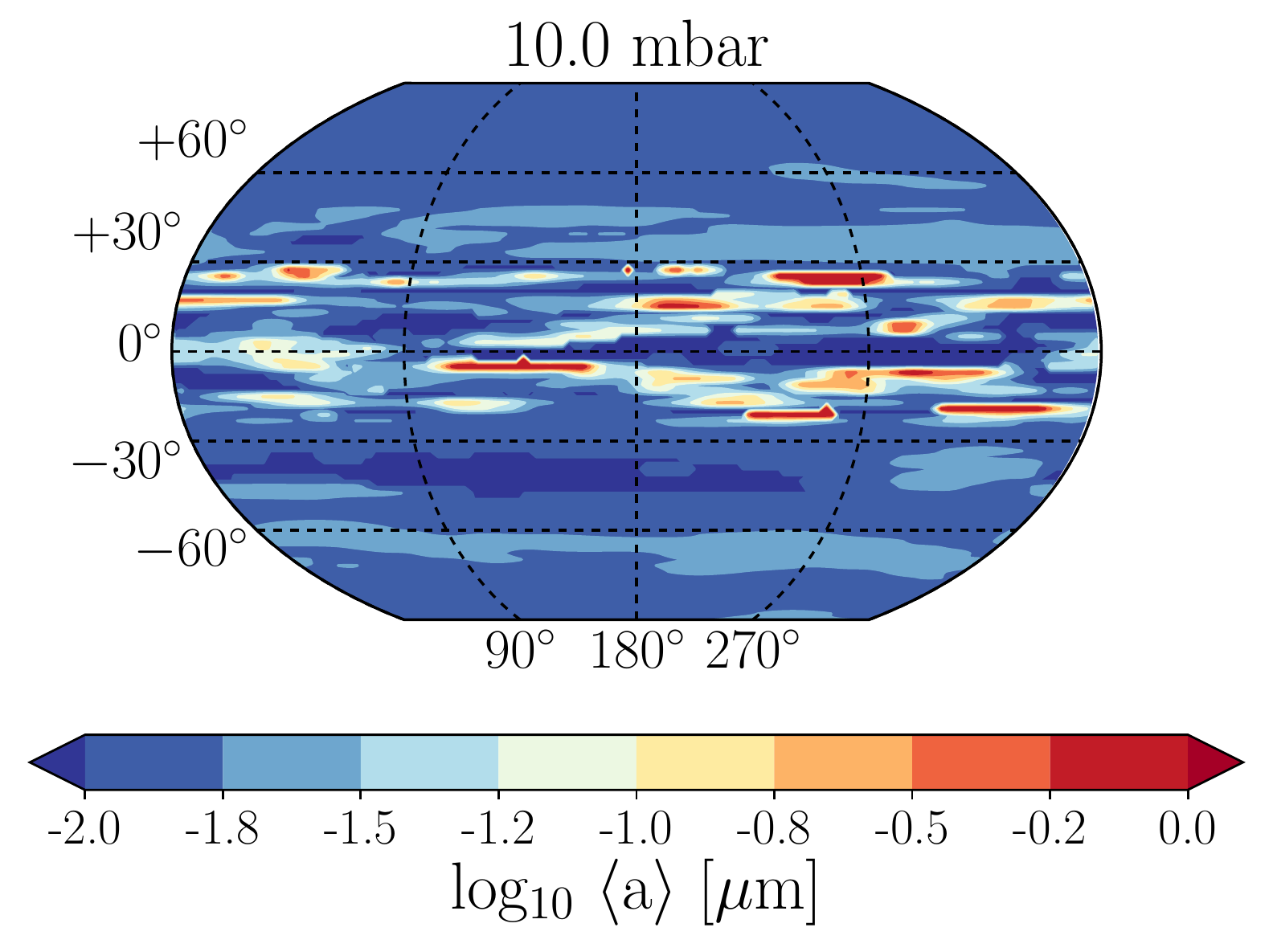}
\endminipage
\caption{Maps of 3D GCM models with cloud formation for HD189733b (left; \citealt{2016A&A...594A..48L}) and for HD\,209458b (right, \citealt{2018arXiv180300226L}). {\bf Top:} gas temperature map, {\bf Centre:} cloud particle number density, {\bf Bottom:} mean cloud particle radii. Both hot Jupiter atmospheres were modelled with different codes such that their latitude grid is offset by 180$^o$ to each other. The point of strongest heating (the substellar point) is for both planets at the equator but at the latitude of 0$^o$ for HD\,189733b (left) and at the latitude of 180$^o$ for HD\,209458b (right).}
\label{fig3D}
\end{figure}

In Figure~\ref{figpies}, the effect of  C/O  on the cloud
material distribution is reviewed for distant giant gas planets of two
different effective temperatures (cold vs warm).  For simplification,
1D cumulative global values were adopted for five groups of
materials: metal oxides (SiO[s], SiO$_2$[s], MgO[s], FeO[s]),
silicates (MgSiO$_3$[s], Mg2SiO$_4$[s]), carbon species (C[s], SiC[s],
TiC[s]), high temperature condensates (TiO$_2$[s], Al$_2$O$_3$[s],
CaTiO$_3$[s], Fe[s]), and sulfur species (S[s], MgS[s], FeS[s]).
KCl[s] and S[s] appeared to be negligible (as nucleation species and
as growth species), hence, it will have little impact on the gas
chemistry.  In the solar system case (left columns), metal oxides and
silicates are the volume dominating condensates except in the case of
a carbon-rich, warm exoplanet. Carbon-binding materials never dominate
the cloud particle material in the summarised cases where C/O
changes  from oxygen-rich  (C/O=0.54) to  carbon-rich
(C/O=1.1), except for the warm model for C/O=2.0. Sulfur haze has been suggested to emerge in
cold gas giants ($<$700K, \citealt{2017AJ....153..139G}). An
overabundance of sulfur was confirmed for Uranus (H$_2$S ice
particles detected, see \citealt{2018NatAs...2..364D}), and volcanic
sulfur as trigger for snowball Earth (\citealt{2017GeoRL..44.1938M}).
Figure~\ref{figpies} contains a first summary in how far an increased
sulfur abundance affects the cloud composition in combination with a
varying C/O. The sulfuric species included in this test do not
contribute much to the cloud material volume in the solar sulfur
abundance cases.  Sulfur plays a larger role as part of condensates
in the warmer atmosphere in the chemical transition region of
C/O=0.99\,$\ldots$\,1.1.

\subsection{Summary of take-away points on element abundances in exoplanet atmospheres}
\begin{enumerate}
\item Spectroscopically observed element abundances of exoplanets are altered by cloud formation and photochemistry.
\item No one carbon-to-oxygen ratio characterises an exoplanet atmosphere.
\item No one mineralogical ratio characterises an exoplanet atmosphere.
\item Observed carbon-rich exoplanets maybe camouflaged cloud-forming oxygen-rich exoplanets.
\end{enumerate}

\section{Cloudy weathers on extrasolar planets}
Forecasting weather on Earth had reached a new level of reliability
when high-performance computations were applied to solve the 3D
radiation hydrodynamics, i.e. the equations that forecast the motion of the atmospheric gas including heating and cooling in 3D. Such models are called global circulation models (GCMs).  The modelling results were compared to and combined
with superb observational data on, for example, precipitation with
respect to particle sizes and their numbers at different
locations. Nevertheless, the Intergovernmental Panel on Climate Change
(IPCC\footnote{http://www.ipcc.ch/}) has identified the effect of
aerosols and precursors as carrying the largest uncertainty among
the drivers for climate change on Earth. No such constraints are
available for exoplanets yet. Seen from Earth, evidence for
weather-like variability on extrasolar planets is provided by phase
curves that show hot spots leading or trailing the point of highest
stellar flux input (substellar point). Nine out of 11 planets show a
4.5$\mu$m--phase-curve  consistent with an east-ward hotspot off-set
(giant-gas planets: HD189744b -- \cite{2007Natur.447..183K}, HD209458b
-- \cite{2014ApJ...790...53Z}, WASP-12b -- \cite{2012ApJ...747...82C},
WASP-14b -- \cite{2015ApJ...811..122W}, WASP19b \& HAT-P-7b --
\cite{2016ApJ...823..122W}, WASP-18b -- \cite{2013MNRAS.428.2645M},
WASP-43b -- \cite{2017AJ....154..232A}; super-Earth: 55 Cnc e --
\cite{2017ApJ...849L...5K}).  The hot Jupiter Kepler-7b\footnote{Kepler is a space telescope built to discover extrasolar planets: \url{https://www.nasa.gov/mission_pages/kepler}.} was the first
planet  detected which had its hottest atmospheric spot off-set to the West (instead to the East; compare Fig.~\ref{fig3D}). This observation is  attributed to high-altitude, optically reflective clouds located west-ward off the substellar point
(\citealt{2013ApJ...776L..25D}).  CoRot-2b\footnote{CoRot stands for  Convection, Rotation and planetary Transits: \url{https://corot.cnes.fr/en/COROT/index.htm}.} is the first hot Jupiter
with a west-ward offset of the hot-spot that is suggested to be caused
by a westward wind due to non-synchronous rotations, magnetic effects,
or maybe partial cloud coverage (\citealt{2018NatAs...2..220D}).  The
partial cloud coverage is supported by a featureless dayside emission
spectrum of CoRoT-2b. Applying 3D cloud-free radiation-hydrodynamics
simulation equatorial jets were
established to be the leading cause of such hot-spots, but additional
mechanisms like clouds were required to explain the observed light
curves (\citealt{2017ApJ...835..198K}).

Global circulation models are extremely time consuming, even more so if sophisticated descriptions of microphysical processes like cloud formation and radiation transfer are included.   Systematic studies of dynamic
features of giant gas planets were therefore conducted based on cloud-
and chemistry-free 3D GCMs (e.g. \citealt{2016ApJ...821....9K}) or
with cloud particles as passive opacity source
(e.g. \citealt{2018arXiv180500096P}). \cite{2018MNRAS.473.4672C}
summaries how the tropospheric circulation states change depending on
planetary radius and orbital distance for tidally locked terrestrial
planets: Close-in orbits of 1 day drive equatorial superrotation and
high-latitude jets (e.g. on TRAPPIST-1b\footnote{TRAPPIST stands for Transiting Planets and Planetesimals Small Telescope which is operated at the La Silla Observatory in Chile: \url{https://en.wikipedia.org/wiki/TRAPPIST}.}). The equatorial superrotation
prevails for increasing orbit of $\sim$10 days
(e.g. TRAPIST-1g). Radial flows drive the atmospheric motion for
orbital periods less than 20 days (e.g. GJ667Cf) as the external
heating trough irradiation has decreased drastically.  Similar flow
pattern will appear in hot Jupiters where the equatorial superrotation
prevails for cloud-free 3D GCM's for HAT-P1b\footnote{HAT stands for Hungarian-made Automated Telescope Network: \url{https://hatnet.org}.}, HAT-P-12b, WASP-6b\footnote{WASP stands for Wide Angle Search for Planets: \url{https://wasp-planets.net}.},
WAS-17b, WASP-19b, WASP-31b, WASP-39b, but also for HD\,189733b and
HD\,209458b (\citealt{2016ApJ...821....9K}). It, however, remains to
be seen in how far different numerical approaches of treating the full
3D radiation hydrodynamics
(\citealt{2013MNRAS.435.3159D,2014A&A...561A...1M,2016ApJ...829..115M})
and variations of 2D+1 hydrodynamics (1 dimension treated in
hydrostatic equilibrium;
\citealt{2013A&A...558A..91P,2017ApJ...835..198K,2018MNRAS.473.4672C})
and additional processes (cloud formation, magnetic coupling)
influence the emergence of superrotation (i.e. extremely fast moving atmospheric gas).

The time-consumption increases substantially if cloud formation and
gas-phase chemistry are included as consistent part of the 3D
radiation-hydrodynamics simulations for exoplanets. Only two planets
have therefore been models so far (HD\,189733b --
\citealt{2016A&A...594A..48L} and HD\,209458b --
\citealt{2018arXiv180300226L}), and no long-term studies were
conducted so far. It was shown that cloud particles of different sizes
and compositions are distributed throughout the whole computational
domain of the modelled atmosphere. Hence, both planets' atmospheres
are affected by cloud formation. Figure~\ref{fig3D} (left:
HD\,189733b, right: HD\,209458b) shows the 3D maps for the gas
temperature (T$_{\rm gas}$ [K]), the number density of cloud particles
($n_{\rm d}$ [cm$^{-3}$]) and mean cloud particle radii ($\langle
a\rangle$ [$\mu$m]) at the pressure level of $\approx 10^{-2}$
bar. Both planets show an off-set of the hottest part of the
atmosphere to the East and the temperature maps demonstrate that the
hydrodynamic transport (wind) is faster than the radiative cooling of
the atmospheric gas. This has strong implications for the cloud
distribution which is most obvious for HD\,189733b at this pressure
level. The cloud particle size map correlated well with  the temperature map in that
the particles are the smallest were the atmospheric gas is the
hottest. Consequently, the cloud particle size maps the east-ward off
set of the temperature. The northern and southern hemispheres on
HD\,189733b contain larger cloud particles than the equatorial
jet-stream region.  For HD\,290458b, the equatorial region shows spots
with large cloud particles whereas the northern and southern
hemisphere are populated with small cloud particles. This is a result
of more cloud particles being present in the equatorial region in
HD\,189733b and less in the hemispheres, and less cloud particles
being present in the equatorial region in HD\,209458b and more in the
hemispheres at this pressure level.  Figure~\ref{fig3D} further shows
how different the two giant-gas planets are regarding their weather
appearance.  HD\,209458b forms three distinct cloud bands, the
retrograde moving northern and southern hemisphere and the prograde
moving equatorial belt. The wave-flow pattern in the equatorial belt
does imprint on the cloud particle sizes appearing as patchy
distribution in Fig.~\ref{fig3D} also in these low-pressure
regions. HD\,189733b has a distinct cloud distribution in the hot-spot
area, in the northern and southern hemispheres and in the equatorial
region. This pattern prevails higher up in the atmosphere where the gas pressure is lowest.

Both planets, HD\,189733b and HD\,209458b, were classified as hot
Jupiters but small differences like in planetary mass and planetary radius (M$_{\rm
  HD189b}\sim 1.3$M$_{\rm J}$, M$_{\rm HD209b}\sim 0.69$M$_{\rm J}$;
R$_{\rm HD189b}\sim 1.138$R$_{\rm J}$, M$_{\rm HD209b}\sim
1.38$R$_{\rm J}$) affect the pressure  stratification of the
atmosphere. The amount of external energy input is different due to
differences in the semi-major axis ($a_{\rm HD189b}\sim 0.031$AU,
$a_{\rm HD209b}\sim 0.047$AU; i.e. the distance to the host star). Both effects combined cause substantial
differences in the local gas densities and gas temperatures which
strongly influence the local gas chemistry, and hence, the local
cloud formation processes. A discussion comparing the cloud properties
of both planets in 1D is provided in (\citealt{2016MNRAS.460..855H}).

\begin{figure}
\vspace*{-0cm}
\hspace*{-0cm}\includegraphics[width=12cm]{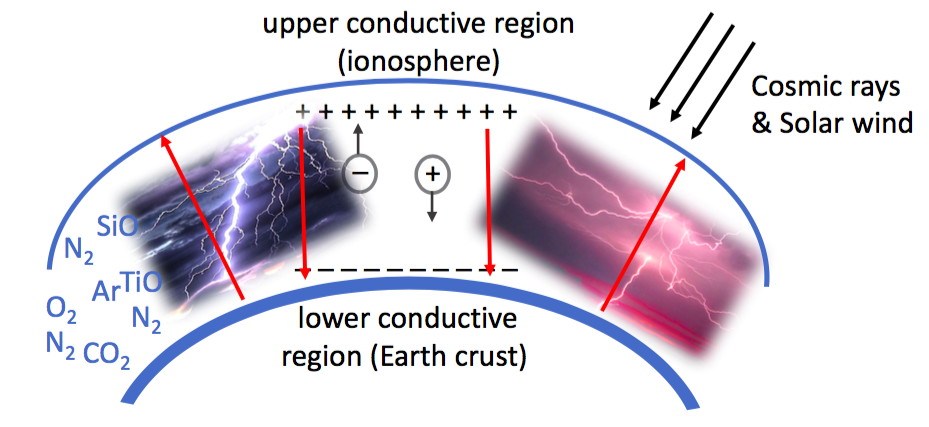}\\*[-0cm]
\caption{The global electric circuit (red arrows) forming in exoplanet
  atmospheres due to lightning activities in exoplanet clouds, the
  formation of an ionosphere through external irradiation (Cosmic
  rays, solar wind) and a inner, conductive boundary.}
  \label{figGEC}
\end{figure}

\subsection{Lightning on exoplanet and the extrasolar global electric circuit}

The CLOUD experiment at CERN\footnote{\url{https://home.cern/about/experiments/cloud}} demonstrated (though only for a very
limited temperature range) that cloud formation expands into
higher atmospheric regions on Earth due to cosmic rays open chemical
channels to form condensation seeds from ionised gas species. The
process becomes inefficient with increasing density, hence
deeper inside the atmosphere. Similar effect would occur for
exoplanets where cosmic rays, stellar and interstellar radiation
affects the gas ionisation in the upper atmosphere such that an
ionosphere may form
(e.g. \citealt{2013ApJ...774..108R,2014IJAsB..13..173R,
  2018arXiv180409054R}).  The ionisation of the gas above the cloud
cause a current which  ionises the upper cloud layers like on
Earth, and the atmosphere will become electrified.  Inside the cloud,
turbulence-driven collisional processes cause triboelectric charging
(see Sect. 2.1 in \citealt{2016SGeo...37..705H}) as the atmosphere is
very dynamic also on small scales
(\citealt{2004A&A...423..657H,2011ApJ...737...38H}).  Gravitational
settling establishes a vertical, large-scale charge separation (see Fig.~\ref{figGEC}). The
advection of aerosols that closely follow the gas motion due to frictionally coupling  may
cause similar effects in horizontal directions. In consequence, an
electrostatic potential difference builds up in various parts of the
atmosphere. This potential can discharge in form of leaking currents
or in form of intra-cloud lightning. Leaking currents ('fair-weather
current') will only occur in cloud-free (parts of) atmospheres, and
lightning only in cloudy (parts of) atmospheres.  Lightning has been
detected on many of the cloudy solar system planets at optical and
radio wavelengths, but a good statistical sampling of the lightning
occurrence rate is only available for Earth
(\citealt{2016MNRAS.461.3927H}). Lightning occurrence rates differ for
oceans and continents on Earth, and hence, may provide a probe for
 different exoplanet environments.

On Earth, lightning plays a major role for feeding what is called a
'global electric circuit' by its return current. The Global Electric Circuit is  a current system that
affects the Earth weather systematically on large and on global scales. The criteria for an
extrasolar global electric circuit include not only
global charge-separation processes and current flows, but also
globally prevalent upper and lower conductive regions (Sect. 3.3 in
\citealt{2016SGeo...37..705H}; see Fig.~\ref{figGEC}). The upper conductivity region is the
ionosphere, and the lower conductivity region will be a rocky planet's
crust or the thermally ionised inner, hot part of a giant gas planet's
atmosphere. It will remain to be seen which role a global electric
circuit has in extrasolar planets which so far have been observed to
possess very dynamical atmospheres with clouds. Maybe electric circuits will
therefore occur on more local, smaller scales in giant exoplanets
compared to smaller, Earth-like planets with less dramatic
hydrodynamics.

In addition to lightning being one of the key elements of a global
electric circuit, lightning dramatically affects the local chemistry
by converting a cold gas into a plasma, hence, by changing its
thermodynamic state from a cold ($\sim$1000K) into a very hot gas
($\sim$30.000K). Possible extrasolar lightning tracer are chemical
species like HCN and C$_2$H
(\citealt{2016ApJS..224....9R,2016MNRAS.461.1222H,2017MNRAS.470..187A}),
or spectral peculiarities like lightning modulated coherent cyclotron
emission coming from electrons that are accelerated by the planet's magnetic field (\citealt{2016MNRAS.458.1041V,2017arXiv171003004H}) similar
to the lightning modulated cosmic ray air showers recently discovered
on Earth (\citealt{2015PhRvL.114p5001S, 2017PhRvD..95h3004T}).
Moreover, strong large-scale magnetic fields like on Saturn (or Brown
Dwarfs) enhance lightning-induced electric fields producing strong
transient optical emission as demonstrated for Earth
(\citealt{2017JGRA..122.7636P}). As lightning requires the presence of
clouds, lightning indicators are formidable tool to prove the presence
of clouds in exoplanet atmosphere as well as gaining insight into the
atmospheric dynamics. If electric circuit systems form on global
scales, it will provide a mechanisms to link atmospheric processes 
 globally.

\subsection{Summary of take-away points in 3D extrasolar clouds}
\begin{enumerate}
\item  Horizontal advection of gas and cloud particles leads to zonal  pattern like bending and hot-spots on exoplanets, similar to Jupiter.
\item Temporary effects on cloud particle distribution cause a time-variable opacity leading to time-dependent thermal emissions and spectral features.
\item Exoplanet clouds can cause lightning which affects the local gas chemistry.
  \item Exoplanet atmospheres may establish a global electric circuit
    that can link atmospheric process over large distances non-locally.
\end{enumerate}




\section{Future issues}
With many satellites and instruments becoming available to analyse
exoplanet atmospheres, cloud formation has turned into a growing
research area that combines the need for extending laboratory work
(e.g. \citealt{0004-637X-527-1-395,2013A&A...554A..12G,2014ApJ...780..180S,2018arXiv180510488H})
to more extreme conditions with that for a fundamental understanding
of phase-transition/cluster chemistry
(e.g. \citealt{2000JPhB...33.3417J,2002CPL...363..145P,2015A&A...575A..11L,2017A&A...608A..55D,2017ApJ...840..117G}),
the exchange with solar system research
(\citealt{2016SGeo...37..705H}) and complex modelling. While first
fully consistent atmosphere models for extrasolar planets have emerged
in 1D and in 3D, approximations are still applied in 1D retrieval
methods and in most 3D simulations in order to limit the demand on
computing times. Such approximations do impact the accuracy of the
local temperature which in turn affects the local chemistry which
determines the spectral appearance. Similar arguments hold for the
treatment of the radiative transfer problem, but here the stellar
literature holds a reservoir (e.g. \citealt{2008PhST..133a4041G,2017MNRAS.469..841H}) to build on  well established
experiences in 1D and in 3D set-ups of varying, wavelength-dependent
optical depths.

Future observations will most likely continue to demonstrate the large
diversity of extrasolar planets of which non resembles any solar
system planet exactly. Venus, Earth and Mars demonstrate how seemingly small differences cause planets to evolve into very different states. The most pressing future issue is therefore to enable
complex, theoretical modelling in tandem with elaborate machine learning tools in order to combine the limited number
of observations per exoplanet such that we are able to make predictions
about the missing information like time-evolution, long-term
behaviour (climate) and maybe even habitability on cosmological scales.

\section*{DISCLOSURE STATEMENT}
If the authors have noting to disclose, the following statement will
be used: The authors are not aware of any affiliations, memberships,
funding, or financial holdings that might be perceived as affecting
the objectivity of this review.

\section*{ACKNOWLEDGMENTS}
David Gobrecht is thanked for providing the (Al$_2$O$_3$)$_{\rm N}$
cluster structures. Jasmina Blecic, Ian Boutle, and  Ludmila Carone are thanks for valuable discussions
on the manuscript. Stefan Lines and Graham Lee are thanked for providing
maps of 3D GCM results.

%

\bibliographystyle{ar-style1.bst}
\bibliography{bib}

\begin{thebibliography}{}
\expandafter\ifx\csname natexlab\endcsname\relax\def\natexlab#1{#1}\fi

\bibitem[{Airapetian} et~al.(2016){Airapetian}, {Glocer}, {Gronoff},
  {H{\'e}brard} \& {Danchi}]{2016NatGe...9..452A}
{Airapetian} VS, {Glocer} A, {Gronoff} G, {H{\'e}brard} E, {Danchi} W. 2016.
{Prebiotic chemistry and atmospheric warming of early Earth by an active young
  Sun}.
\textit{Nature Geoscience} 9:452--455

\bibitem[{Angelo} \& {Hu}(2017)]{2017AJ....154..232A}
{Angelo} I, {Hu} R. 2017.
{A Case for an Atmosphere on Super-Earth 55 Cancri e}.
\textit{\aj} 154:232

\bibitem[{Apai} et~al.(2017){Apai}, {Karalidi}, {Marley}, {Yang}, {Flateau}
  et~al.]{2017Sci...357..683A}
{Apai} D, {Karalidi} T, {Marley} MS, {Yang} H, {Flateau} D, et~al. 2017.
{Zones, spots, and planetary-scale waves beating in brown dwarf atmospheres}.
\textit{Science} 357:683--687

\bibitem[{Apai} et~al.(2013){Apai}, {Radigan}, {Buenzli}, {Burrows}, {Reid} \&
  {Jayawardhana}]{2013ApJ...768..121A}
{Apai} D, {Radigan} J, {Buenzli} E, {Burrows} A, {Reid} IN, {Jayawardhana} R.
  2013.
{HST Spectral Mapping of L/T Transition Brown Dwarfs Reveals Cloud Thickness
  Variations}.
\textit{\apj} 768:121

\bibitem[{Arcangeli} et~al.(2018){Arcangeli}, {D{\'e}sert}, {Line}, {Bean},
  {Parmentier} et~al.]{2018ApJ...855L..30A}
{Arcangeli} J, {D{\'e}sert} JM, {Line} MR, {Bean} JL, {Parmentier} V, et~al.
  2018.
{H$^{?}$ Opacity and Water Dissociation in the Dayside Atmosphere of the Very
  Hot Gas Giant WASP-18b}.
\textit{\apjl} 855:L30

\bibitem[{Ardaseva} et~al.(2017){Ardaseva}, {Rimmer}, {Waldmann}, {Rocchetto},
  {Yurchenko} et~al.]{2017MNRAS.470..187A}
{Ardaseva} A, {Rimmer} PB, {Waldmann} I, {Rocchetto} M, {Yurchenko} SN, et~al.
  2017.
{Lightning chemistry on Earth-like exoplanets}.
\textit{\mnras} 470:187--196

\bibitem[{Bilger} et~al.(2013){Bilger}, {Rimmer} \&
  {Helling}]{2013MNRAS.435.1888B}
{Bilger} C, {Rimmer} P, {Helling} C. 2013.
{Small hydrocarbon molecules in cloud-forming brown dwarf and giant gas planet
  atmospheres}.
\textit{\mnras} 435:1888--1903

\bibitem[{Blecic} et~al.(2017){Blecic}, {Dobbs-Dixon} \&
  {Greene}]{2017ApJ...848..127B}
{Blecic} J, {Dobbs-Dixon} I, {Greene} T. 2017.
{The Implications of 3D Thermal Structure on 1D Atmospheric Retrieval}.
\textit{\apj} 848:127

\bibitem[{Carone} et~al.(2018){Carone}, {Keppens}, {Decin} \&
  {Henning}]{2018MNRAS.473.4672C}
{Carone} L, {Keppens} R, {Decin} L, {Henning} T. 2018.
{Stratosphere circulation on tidally locked ExoEarths}.
\textit{\mnras} 473:4672--4685

\bibitem[{Carter} et~al.(2015){Carter}, {Leinhardt}, {Elliott}, {Walter} \&
  {Stewart}]{2015ApJ...813...72C}
{Carter} PJ, {Leinhardt} ZM, {Elliott} T, {Walter} MJ, {Stewart} ST. 2015.
{Compositional Evolution during Rocky Protoplanet Accretion}.
\textit{\apj} 813:72

\bibitem[{Charnay} et~al.(2018){Charnay}, {B{\'e}zard}, {Baudino}, {Bonnefoy},
  {Boccaletti} \& {Galicher}]{2018ApJ...854..172C}
{Charnay} B, {B{\'e}zard} B, {Baudino} JL, {Bonnefoy} M, {Boccaletti} A,
  {Galicher} R. 2018.
{A Self-consistent Cloud Model for Brown Dwarfs and Young Giant Exoplanets:
  Comparison with Photometric and Spectroscopic Observations}.
\textit{\apj} 854:172

\bibitem[{Cowan} et~al.(2012){Cowan}, {Machalek}, {Croll}, {Shekhtman},
  {Burrows} et~al.]{2012ApJ...747...82C}
{Cowan} NB, {Machalek} P, {Croll} B, {Shekhtman} LM, {Burrows} A, et~al. 2012.
{Thermal Phase Variations of WASP-12b: Defying Predictions}.
\textit{\apj} 747:82

\bibitem[{Crida} et~al.(2018){Crida}, {Ligi}, {Dorn} \&
  {Lebreton}]{2018arXiv180407537C}
{Crida} A, {Ligi} R, {Dorn} C, {Lebreton} Y. 2018.
{Mass, radius, and composition of the transiting planet 55 Cnc e : using
  interferometry and correlations}.
\textit{ArXiv e-prints}

\bibitem[{Cridland} et~al.(2016){Cridland}, {Pudritz} \&
  {Alessi}]{2016MNRAS.461.3274C}
{Cridland} AJ, {Pudritz} RE, {Alessi} M. 2016.
{Composition of early planetary atmospheres - I. Connecting disc astrochemistry
  to the formation of planetary atmospheres}.
\textit{\mnras} 461:3274--3295

\bibitem[{Dang} et~al.(2018){Dang}, {Cowan}, {Schwartz}, {Rauscher}, {Zhang}
  et~al.]{2018NatAs...2..220D}
{Dang} L, {Cowan} NB, {Schwartz} JC, {Rauscher} E, {Zhang} M, et~al. 2018.
{Detection of a westward hotspot offset in the atmosphere of hot gas giant
  CoRoT-2b}.
\textit{Nature Astronomy} 2:220--227

\bibitem[{de Pater}(2018)]{2018NatAs...2..364D}
{de Pater} I. 2018.
{Selective enrichment of volatiles confirmed}.
\textit{Nature Astronomy} 2:364--365

\bibitem[{Decin} et~al.(2017){Decin}, {Richards}, {Waters}, {Danilovich},
  {Gobrecht} et~al.]{2017A&A...608A..55D}
{Decin} L, {Richards} AMS, {Waters} LBFM, {Danilovich} T, {Gobrecht} D, et~al.
  2017.
{Study of the aluminium content in AGB winds using ALMA. Indications for the
  presence of gas-phase (Al$_{2}$O$_{3}$)$_{n}$ clusters}.
\textit{\aap} 608:A55

\bibitem[{Demory} et~al.(2013){Demory}, {de Wit}, {Lewis}, {Fortney}, {Zsom}
  et~al.]{2013ApJ...776L..25D}
{Demory} BO, {de Wit} J, {Lewis} N, {Fortney} J, {Zsom} A, et~al. 2013.
{Inference of Inhomogeneous Clouds in an Exoplanet Atmosphere}.
\textit{\apjl} 776:L25

\bibitem[{Diamond-Lowe} et~al.(2018){Diamond-Lowe}, {Berta-Thompson},
  {Charbonneau} \& {Kempton}]{2018arXiv180507328D}
{Diamond-Lowe} H, {Berta-Thompson} Z, {Charbonneau} D, {Kempton} EMR. 2018.
{Ground-based optical transmission spectroscopy of the small, rocky exoplanet
  GJ 1132b}.
\textit{ArXiv e-prints}

\bibitem[{Dobbs-Dixon} \& {Agol}(2013)]{2013MNRAS.435.3159D}
{Dobbs-Dixon} I, {Agol} E. 2013.
{Three-dimensional radiative-hydrodynamical simulations of the highly
  irradiated short-period exoplanet HD 189733b}.
\textit{\mnras} 435:3159--3168

\bibitem[{Dunne} et~al.(2016){Dunne}, {Gordon}, {K{\"u}rten}, {Almeida},
  {Duplissy} et~al.]{2016Sci...354.1119D}
{Dunne} EM, {Gordon} H, {K{\"u}rten} A, {Almeida} J, {Duplissy} J, et~al. 2016.
{Global atmospheric particle formation from CERN CLOUD measurements}.
\textit{Science} 354:1119--1124

\bibitem[{Eistrup} et~al.(2016){Eistrup}, {Walsh} \& {van
  Dishoeck}]{2016A&A...595A..83E}
{Eistrup} C, {Walsh} C, {van Dishoeck} EF. 2016.
{Setting the volatile composition of (exo)planet-building material. Does
  chemical evolution in disk midplanes matter?}
\textit{\aap} 595:A83

\bibitem[{Gadallah} et~al.(2013){Gadallah}, {Mutschke} \&
  {J{\"a}ger}]{2013A&A...554A..12G}
{Gadallah} KAK, {Mutschke} H, {J{\"a}ger} C. 2013.
{Analogs of solid nanoparticles as precursors of aromatic hydrocarbons}.
\textit{\aap} 554:A12

\bibitem[{Gao} et~al.(2017){Gao}, {Marley}, {Zahnle}, {Robinson} \&
  {Lewis}]{2017AJ....153..139G}
{Gao} P, {Marley} MS, {Zahnle} K, {Robinson} TD, {Lewis} NK. 2017.
{Sulfur Hazes in Giant Exoplanet Atmospheres: Impacts on Reflected Light
  Spectra}.
\textit{\aj} 153:139

\bibitem[{Glawe} et~al.(2015){Glawe}, {Schmidt}, {Kerstein} \&
  {Klein}]{2015arXiv150604930G}
{Glawe} C, {Schmidt} H, {Kerstein} AR, {Klein} R. 2015.
{XLES Part I: Introduction to Extended Large Eddy Simulation}.
\textit{ArXiv e-prints}

\bibitem[{Gobrecht} et~al.(2017){Gobrecht}, {Cristallo}, {Piersanti} \&
  {Bromley}]{2017ApJ...840..117G}
{Gobrecht} D, {Cristallo} S, {Piersanti} L, {Bromley} ST. 2017.
{Nucleation of Small Silicon Carbide Dust Clusters in AGB Stars}.
\textit{\apj} 840:117

\bibitem[{Goeres}(1996)]{1996ASPC...96...69G}
{Goeres} A. 1996.
{Chemistry and thermodynamics of the nucleation in R CrB star shells}, In
  \textit{Hydrogen Deficient Stars}, eds. CS~{Jeffery}, U~{Heber}, vol.~96 of
  \textit{Astronomical Society of the Pacific Conference Series}

\bibitem[{Gustafsson}(2008)]{2008PhST..133a4041G}
{Gustafsson} B. 2008.
{An attempt to summarize and conclude}.
\textit{Physica Scripta Volume T} 133:014041

\bibitem[{Gustafsson} et~al.(2008){Gustafsson}, {Edvardsson}, {Eriksson},
  {J{\o}rgensen}, {Nordlund} \& {Plez}]{2008A&A...486..951G}
{Gustafsson} B, {Edvardsson} B, {Eriksson} K, {J{\o}rgensen} UG, {Nordlund}
  {\AA}, {Plez} B. 2008.
{A grid of MARCS model atmospheres for late-type stars. I. Methods and general
  properties}.
\textit{\aap} 486:951--970

\bibitem[{He} et~al.(2018){He}, {Horst}, {Lewis}, {Yu}, {Moses}
  et~al.]{2018arXiv180510488H}
{He} C, {Horst} SM, {Lewis} NK, {Yu} X, {Moses} JI, et~al. 2018.
{Photochemical Haze Formation in the Atmospheres of super-Earths and
  mini-Neptunes}.
\textit{ArXiv e-prints}

\bibitem[{Helling} et~al.(2008{\natexlab{a}}){Helling}, {Ackerman}, {Allard},
  {Dehn}, {Hauschildt} et~al.]{2008MNRAS.391.1854H}
{Helling} C, {Ackerman} A, {Allard} F, {Dehn} M, {Hauschildt} P, et~al.
  2008{\natexlab{a}}.
{A comparison of chemistry and dust cloud formation in ultracool dwarf model
  atmospheres}.
\textit{\mnras} 391:1854--1873

\bibitem[{Helling} \& {Fomins}(2013)]{2013RSPTA.37110581H}
{Helling} C, {Fomins} A. 2013.
{Modelling the formation of atmospheric dust in brown dwarfs and planetary
  atmospheres}.
\textit{Philosophical Transactions of the Royal Society of London Series A}
  371:20110581--20110581

\bibitem[{Helling} et~al.(2016{\natexlab{a}}){Helling}, {Harrison}, {Honary},
  {Diver}, {Aplin} et~al.]{2016SGeo...37..705H}
{Helling} C, {Harrison} RG, {Honary} F, {Diver} DA, {Aplin} K, et~al.
  2016{\natexlab{a}}.
{Atmospheric Electrification in Dusty, Reactive Gases in the Solar System and
  Beyond}.
\textit{Surveys in Geophysics} 37:705--756

\bibitem[{Helling} et~al.(2011){Helling}, {Jardine} \&
  {Mokler}]{2011ApJ...737...38H}
{Helling} C, {Jardine} M, {Mokler} F. 2011.
{Ionization in Atmospheres of Brown Dwarfs and Extrasolar Planets. II.
  Dust-induced Collisional Ionization}.
\textit{\apj} 737:38

\bibitem[{Helling} et~al.(2004){Helling}, {Klein}, {Woitke}, {Nowak} \&
  {Sedlmayr}]{2004A&A...423..657H}
{Helling} C, {Klein} R, {Woitke} P, {Nowak} U, {Sedlmayr} E. 2004.
{Dust in brown dwarfs. IV. Dust formation and driven turbulence on mesoscopic
  scales}.
\textit{\aap} 423:657--675

\bibitem[{Helling} et~al.(2016{\natexlab{b}}){Helling}, {Lee}, {Dobbs-Dixon},
  {Mayne}, {Amundsen} et~al.]{2016MNRAS.460..855H}
{Helling} C, {Lee} G, {Dobbs-Dixon} I, {Mayne} N, {Amundsen} DS, et~al.
  2016{\natexlab{b}}.
{The mineral clouds on HD 209458b and HD 189733b}.
\textit{\mnras} 460:855--883

\bibitem[{Helling} et~al.(2016{\natexlab{c}}){Helling}, {Rimmer},
  {Rodriguez-Barrera}, {Wood}, {Robertson} \& {Stark}]{2016PPCF...58g4003H}
{Helling} C, {Rimmer} PB, {Rodriguez-Barrera} IM, {Wood} K, {Robertson} GB,
  {Stark} CR. 2016{\natexlab{c}}.
{Ionisation and discharge in cloud-forming atmospheres of brown dwarfs and
  extrasolar planets}.
\textit{Plasma Physics and Controlled Fusion} 58:074003

\bibitem[{Helling} \& {Vorgul}(2017)]{2017arXiv171003004H}
{Helling} C, {Vorgul} I. 2017.
{Insight into atmospheres of extrasolar planets through plasma processes}.
\textit{ArXiv e-prints}

\bibitem[{Helling} et~al.(2014){Helling}, {Woitke}, {Rimmer}, {Kamp}, {Thi} \&
  {Meijerink}]{2014Life....4..142H}
{Helling} C, {Woitke} P, {Rimmer} PB, {Kamp} I, {Thi} WF, {Meijerink} R. 2014.
{Disk Evolution, Element Abundances and Cloud Properties of Young Gas Giant
  Planets}.
\textit{Life} 4

\bibitem[{Helling} et~al.(2008{\natexlab{b}}){Helling}, {Woitke} \&
  {Thi}]{2008A&A...485..547H}
{Helling} C, {Woitke} P, {Thi} WF. 2008{\natexlab{b}}.
{Dust in brown dwarfs and extra-solar planets. I. Chemical composition and
  spectral appearance of quasi-static cloud layers}.
\textit{\aap} 485:547--560

\bibitem[{Hellmuth}(2006)]{2006ACP.....6.4175H}
{Hellmuth} O. 2006.
{Columnar modelling of nucleation burst evolution in the convective boundary
  layer - first results from a feasibility study Part I: Modelling approach}.
\textit{Atmospheric Chemistry \& Physics} 6:4175--4214

\bibitem[{Hodos{\'a}n} et~al.(2016{\natexlab{a}}){Hodos{\'a}n}, {Helling},
  {Asensio-Torres}, {Vorgul} \& {Rimmer}]{2016MNRAS.461.3927H}
{Hodos{\'a}n} G, {Helling} C, {Asensio-Torres} R, {Vorgul} I, {Rimmer} PB.
  2016{\natexlab{a}}.
{Lightning climatology of exoplanets and brown dwarfs guided by Solar system
  data}.
\textit{\mnras} 461:3927--3947

\bibitem[{Hodos{\'a}n} et~al.(2016{\natexlab{b}}){Hodos{\'a}n}, {Rimmer} \&
  {Helling}]{2016MNRAS.461.1222H}
{Hodos{\'a}n} G, {Rimmer} PB, {Helling} C. 2016{\natexlab{b}}.
{Is lightning a possible source of the radio emission on HAT-P-11b?}
\textit{\mnras} 461:1222--1226

\bibitem[{Hubeny}(2017)]{2017MNRAS.469..841H}
{Hubeny} I. 2017.
{Model atmospheres of sub-stellar mass objects}.
\textit{\mnras} 469:841--869

\bibitem[{Jeong} et~al.(2000){Jeong}, {Chang}, {Sedlmayr} \&
  {S{\"u}lzle}]{2000JPhB...33.3417J}
{Jeong} KS, {Chang} C, {Sedlmayr} E, {S{\"u}lzle} D. 2000.
{Electronic structure investigation of neutral titanium oxide molecules
  Ti$_{x}$O$_{y}$}.
\textit{Journal of Physics B Atomic Molecular Physics} 33:3417--3430

\bibitem[{Kataria} et~al.(2016){Kataria}, {Sing}, {Lewis}, {Visscher},
  {Showman} et~al.]{2016ApJ...821....9K}
{Kataria} T, {Sing} DK, {Lewis} NK, {Visscher} C, {Showman} AP, et~al. 2016.
{The Atmospheric Circulation of a Nine-hot-Jupiter Sample: Probing Circulation
  and Chemistry over a Wide Phase Space}.
\textit{\apj} 821:9

\bibitem[{Kawashima} \& {Ikoma}(2018)]{2018ApJ...853....7K}
{Kawashima} Y, {Ikoma} M. 2018.
{Theoretical Transmission Spectra of Exoplanet Atmospheres with Hydrocarbon
  Haze: Effect of Creation, Growth, and Settling of Haze Particles. I. Model
  Description and First Results}.
\textit{\apj} 853:7

\bibitem[{Keating} \& {Cowan}(2017)]{2017ApJ...849L...5K}
{Keating} D, {Cowan} NB. 2017.
{Revisiting the Energy Budget of WASP-43b: Enhanced Day-Night Heat Transport}.
\textit{\apjl} 849:L5

\bibitem[{Knutson} et~al.(2007){Knutson}, {Charbonneau}, {Allen}, {Fortney},
  {Agol} et~al.]{2007Natur.447..183K}
{Knutson} HA, {Charbonneau} D, {Allen} LE, {Fortney} JJ, {Agol} E, et~al. 2007.
{A map of the day-night contrast of the extrasolar planet HD 189733b}.
\textit{\nat} 447:183--186

\bibitem[{Komacek} et~al.(2017){Komacek}, {Showman} \&
  {Tan}]{2017ApJ...835..198K}
{Komacek} TD, {Showman} AP, {Tan} X. 2017.
{Atmospheric Circulation of Hot Jupiters: Dayside-Nightside Temperature
  Differences. II. Comparison with Observations}.
\textit{\apj} 835:198

\bibitem[{Kreidberg} et~al.(2014){Kreidberg}, {Bean}, {D{\'e}sert}, {Benneke},
  {Deming} et~al.]{2014Natur.505...69K}
{Kreidberg} L, {Bean} JL, {D{\'e}sert} JM, {Benneke} B, {Deming} D, et~al.
  2014.
{Clouds in the atmosphere of the super-Earth exoplanet GJ1214b}.
\textit{\nat} 505:69--72

\bibitem[{Kr\"uger} \& {Sedlmayr}(1997)]{1997A&A...321..557K}
{Kr\"uger} D, {Sedlmayr} E. 1997.
{Two-fluid models for stationary dust driven winds. II. The grain size
  distribution in consideration of drift.}
\textit{\aap} 321:557--567

\bibitem[{Lee} et~al.(2016){Lee}, {Dobbs-Dixon}, {Helling}, {Bognar} \&
  {Woitke}]{2016A&A...594A..48L}
{Lee} G, {Dobbs-Dixon} I, {Helling} C, {Bognar} K, {Woitke} P. 2016.
{Dynamic mineral clouds on HD 189733b. I. 3D RHD with kinetic, non-equilibrium
  cloud formation}.
\textit{\aap} 594:A48

\bibitem[{Lee} et~al.(2015){Lee}, {Helling}, {Giles} \&
  {Bromley}]{2015A&A...575A..11L}
{Lee} G, {Helling} C, {Giles} H, {Bromley} ST. 2015.
{Dust in brown dwarfs and extra-solar planets. IV. Assessing TiO$_{2}$ and SiO
  nucleation for cloud formation modelling}.
\textit{\aap} 575:A11

\bibitem[{Lines} et~al.(2018){Lines}, {Mayne}, {Boutle}, {Manners}, {Lee}
  et~al.]{2018arXiv180300226L}
{Lines} S, {Mayne} NJ, {Boutle} IA, {Manners} J, {Lee} GKH, et~al. 2018.
{Simulating the cloudy atmospheres of HD 209458 b and HD 189733 b with the 3D
  Met Office Unified Model}.
\textit{ArXiv e-prints}

\bibitem[{Lothringer} et~al.(2018){Lothringer}, {Barman} \&
  {Koskinen}]{2018arXiv180500038L}
{Lothringer} JD, {Barman} T, {Koskinen} T. 2018.
{Extremely Irradiated Hot Jupiters: Non-Oxide Inversions, H- Opacity, and
  Thermal Dissociation of Molecules}.
\textit{ArXiv e-prints}

\bibitem[{M{\"a}{\"a}tt{\"a}nen} et~al.(2010){M{\"a}{\"a}tt{\"a}nen},
  {Montmessin}, {Gondet}, {Scholten}, {Hoffmann} et~al.]{2010Icar..209..452M}
{M{\"a}{\"a}tt{\"a}nen} A, {Montmessin} F, {Gondet} B, {Scholten} F, {Hoffmann}
  H, et~al. 2010.
{Mapping the mesospheric CO $_{2}$ clouds on Mars: MEx/OMEGA and MEx/HRSC
  observations and challenges for atmospheric models}.
\textit{\icarus} 209:452--469

\bibitem[{Macdonald} \& {Wordsworth}(2017)]{2017GeoRL..44.1938M}
{Macdonald} FA, {Wordsworth} R. 2017.
{Initiation of Snowball Earth with volcanic sulfur aerosol emissions}.
\textit{\grl} 44:1938--1946

\bibitem[{Madhusudhan} et~al.(2017){Madhusudhan}, {Bitsch}, {Johansen} \&
  {Eriksson}]{2017MNRAS.469.4102M}
{Madhusudhan} N, {Bitsch} B, {Johansen} A, {Eriksson} L. 2017.
{Atmospheric signatures of giant exoplanet formation by pebble accretion}.
\textit{\mnras} 469:4102--4115

\bibitem[{Mahapatra} et~al.(2017){Mahapatra}, {Helling} \&
  {Miguel}]{2017MNRAS.472..447M}
{Mahapatra} G, {Helling} C, {Miguel} Y. 2017.
{Cloud formation in metal-rich atmospheres of hot super-Earths like 55 Cnc e
  and CoRoT7b}.
\textit{\mnras} 472:447--464

\bibitem[{Marley} \& {Robinson}(2015)]{2015ARA&A..53..279M}
{Marley} MS, {Robinson} TD. 2015.
{On the Cool Side: Modeling the Atmospheres of Brown Dwarfs and Giant Planets}.
\textit{\araa} 53:279--323

\bibitem[{Maxted} et~al.(2013){Maxted}, {Anderson}, {Doyle}, {Gillon},
  {Harrington} et~al.]{2013MNRAS.428.2645M}
{Maxted} PFL, {Anderson} DR, {Doyle} AP, {Gillon} M, {Harrington} J, et~al.
  2013.
{Spitzer 3.6 and 4.5 {$\mu$}m full-orbit light curves of WASP-18}.
\textit{\mnras} 428:2645--2660

\bibitem[{Mayne} et~al.(2014){Mayne}, {Baraffe}, {Acreman}, {Smith}, {Browning}
  et~al.]{2014A&A...561A...1M}
{Mayne} NJ, {Baraffe} I, {Acreman} DM, {Smith} C, {Browning} MK, et~al. 2014.
{The unified model, a fully-compressible, non-hydrostatic, deep atmosphere
  global circulation model, applied to hot Jupiters. ENDGame for a HD 209458b
  test case}.
\textit{\aap} 561:A1

\bibitem[{Mendon{\c c}a} et~al.(2016){Mendon{\c c}a}, {Grimm}, {Grosheintz} \&
  {Heng}]{2016ApJ...829..115M}
{Mendon{\c c}a} JM, {Grimm} SL, {Grosheintz} L, {Heng} K. 2016.
{THOR: A New and Flexible Global Circulation Model to Explore Planetary
  Atmospheres}.
\textit{\apj} 829:115

\bibitem[{Moses} et~al.(2000){Moses}, {B{\'e}zard}, {Lellouch}, {Gladstone},
  {Feuchtgruber} \& {Allen}]{2000Icar..143..244M}
{Moses} JI, {B{\'e}zard} B, {Lellouch} E, {Gladstone} GR, {Feuchtgruber} H,
  {Allen} M. 2000.
{Photochemistry of Saturn's Atmosphere. I. Hydrocarbon Chemistry and
  Comparisons with ISO Observations}.
\textit{\icarus} 143:244--298

\bibitem[{Moses} et~al.(2016){Moses}, {Marley}, {Zahnle}, {Line}, {Fortney}
  et~al.]{2016ApJ...829...66M}
{Moses} JI, {Marley} MS, {Zahnle} K, {Line} MR, {Fortney} JJ, et~al. 2016.
{On the Composition of Young, Directly Imaged Giant Planets}.
\textit{\apj} 829:66

\bibitem[{Parmentier} et~al.(2018){Parmentier}, {Line}, {Bean}, {Mansfield},
  {Kreidberg} et~al.]{2018arXiv180500096P}
{Parmentier} V, {Line} MR, {Bean} JL, {Mansfield} M, {Kreidberg} L, et~al.
  2018.
{From thermal dissociation to condensation in the atmospheres of ultra hot
  Jupiters: WASP-121b in context}.
\textit{ArXiv e-prints}

\bibitem[{Parmentier} et~al.(2013){Parmentier}, {Showman} \&
  {Lian}]{2013A&A...558A..91P}
{Parmentier} V, {Showman} AP, {Lian} Y. 2013.
{3D mixing in hot Jupiters atmospheres. I. Application to the day/night cold
  trap in HD 209458b}.
\textit{\aap} 558:A91

\bibitem[{Patzer} et~al.(2002){Patzer}, {Chang}, {John}, {Bolick} \&
  {S{\"u}lzle}]{2002CPL...363..145P}
{Patzer} ABC, {Chang} C, {John} M, {Bolick} U, {S{\"u}lzle} D. 2002.
{Theoretical study of stationary points of the MgSiO $_{3}$ molecule}.
\textit{Chemical Physics Letters} 363:145--151

\bibitem[{P{\'e}rez-Invern{\'o}n} et~al.(2017){P{\'e}rez-Invern{\'o}n}, {Luque}
  \& {Gordillo-V{\'a}zquez}]{2017JGRA..122.7636P}
{P{\'e}rez-Invern{\'o}n} FJ, {Luque} A, {Gordillo-V{\'a}zquez} FJ. 2017.
{Three-dimensional modeling of lightning-induced electromagnetic pulses on
  Venus, Jupiter, and Saturn}.
\textit{Journal of Geophysical Research (Space Physics)} 122:7636--7653

\bibitem[{Pinhas} et~al.(2016){Pinhas}, {Madhusudhan} \&
  {Clarke}]{2016MNRAS.463.4516P}
{Pinhas} A, {Madhusudhan} N, {Clarke} C. 2016.
{Efficiency of planetesimal ablation in giant planetary envelopes}.
\textit{\mnras} 463:4516--4532

\bibitem[{Plane} et~al.(2018){Plane}, {Flynn}, {M{\"a}{\"a}tt{\"a}nen},
  {Moores}, {Poppe} et~al.]{2018SSRv..214...23P}
{Plane} JMC, {Flynn} GJ, {M{\"a}{\"a}tt{\"a}nen} A, {Moores} JE, {Poppe} AR,
  et~al. 2018.
{Impacts of Cosmic Dust on Planetary Atmospheres and Surfaces}.
\textit{\ssr} 214:23

\bibitem[{Pont} et~al.(2013){Pont}, {Sing}, {Gibson}, {Aigrain}, {Henry} \&
  {Husnoo}]{2013MNRAS.432.2917P}
{Pont} F, {Sing} DK, {Gibson} NP, {Aigrain} S, {Henry} G, {Husnoo} N. 2013.
{The prevalence of dust on the exoplanet HD 189733b from Hubble and Spitzer
  observations}.
\textit{\mnras} 432:2917--2944

\bibitem[{Powell} et~al.(2018){Powell}, {Zhang}, {Gao} \&
  {Parmentier}]{2018arXiv180501468P}
{Powell} D, {Zhang} X, {Gao} P, {Parmentier} V. 2018.
{Formation of Silicate and Titanium Clouds on Hot Jupiters}.
\textit{ArXiv e-prints}

\bibitem[Rietmeijer et~al.(1999)Rietmeijer, III \& Karner]{0004-637X-527-1-395}
Rietmeijer FJM, III JAN, Karner JM. 1999.
Metastable eutectic condensation in a mg-fe-sio-h2-o2 vapor: Analogs to
  circumstellar dust.
\textit{The Astrophysical Journal} 527:395

\bibitem[{Rimmer} \& {Helling}(2013)]{2013ApJ...774..108R}
{Rimmer} PB, {Helling} C. 2013.
{Ionization in Atmospheres of Brown Dwarfs and Extrasolar Planets. IV. The
  Effect of Cosmic Rays}.
\textit{\apj} 774:108

\bibitem[{Rimmer} \& {Helling}(2016)]{2016ApJS..224....9R}
{Rimmer} PB, {Helling} C. 2016.
{A Chemical Kinetics Network for Lightning and Life in Planetary Atmospheres}.
\textit{\apjs} 224:9

\bibitem[{Rimmer} et~al.(2014){Rimmer}, {Helling} \&
  {Bilger}]{2014IJAsB..13..173R}
{Rimmer} PB, {Helling} C, {Bilger} C. 2014.
{The influence of galactic cosmic rays on ion-neutral hydrocarbon chemistry in
  the upper atmospheres of free-floating exoplanets}.
\textit{International Journal of Astrobiology} 13:173--181

\bibitem[{Rodriguez-Barrera} et~al.(2018){Rodriguez-Barrera}, {Helling} \&
  {Wood}]{2018arXiv180409054R}
{Rodriguez-Barrera} MI, {Helling} C, {Wood} K. 2018.
{Environmental effects on the ionization of brown dwarf atmospheres}.
\textit{ArXiv e-prints}

\bibitem[{Sabri} et~al.(2014){Sabri}, {Gavilan}, {J{\"a}ger}, {Lemaire},
  {Vidali} et~al.]{2014ApJ...780..180S}
{Sabri} T, {Gavilan} L, {J{\"a}ger} C, {Lemaire} JL, {Vidali} G, et~al. 2014.
{Interstellar Silicate Analogs for Grain-surface Reaction Experiments:
  Gas-phase Condensation and Characterization of the Silicate Dust Grains}.
\textit{\apj} 780:180

\bibitem[{Sagan} \& {Khare}(1979)]{1979Natur.277..102S}
{Sagan} C, {Khare} BN. 1979.
{Tholins - Organic chemistry of interstellar grains and gas}.
\textit{\nat} 277:102--107

\bibitem[{Santos} et~al.(2017){Santos}, {Adibekyan}, {Dorn}, {Mordasini},
  {Noack} et~al.]{2017A&A...608A..94S}
{Santos} NC, {Adibekyan} V, {Dorn} C, {Mordasini} C, {Noack} L, et~al. 2017.
{Constraining planet structure and composition from stellar chemistry: trends
  in different stellar populations}.
\textit{\aap} 608:A94

\bibitem[{Schellart} et~al.(2015){Schellart}, {Trinh}, {Buitink}, {Corstanje},
  {Enriquez} et~al.]{2015PhRvL.114p5001S}
{Schellart} P, {Trinh} TNG, {Buitink} S, {Corstanje} A, {Enriquez} JE, et~al.
  2015.
{Probing Atmospheric Electric Fields in Thunderstorms through Radio Emission
  from Cosmic-Ray-Induced Air Showers}.
\textit{Physical Review Letters} 114:165001

\bibitem[{Showman} et~al.(2008){Showman}, {Cooper}, {Fortney} \&
  {Marley}]{2008ApJ...682..559S}
{Showman} AP, {Cooper} CS, {Fortney} JJ, {Marley} MS. 2008.
{Atmospheric Circulation of Hot Jupiters: Three-dimensional Circulation Models
  of HD 209458b and HD 189733b with Simplified Forcing}.
\textit{\apj} 682:559--576

\bibitem[{Smith}(1990)]{1990QJRMS.116..435S}
{Smith} RNB. 1990.
{A scheme for predicting layer clouds and their water content in a general
  circulation model}.
\textit{Quarterly Journal of the Royal Meteorological Society} 116:435--460

\bibitem[{Snellen} et~al.(2013){Snellen}, {de Kok}, {le Poole}, {Brogi} \&
  {Birkby}]{2013ApJ...764..182S}
{Snellen} IAG, {de Kok} RJ, {le Poole} R, {Brogi} M, {Birkby} J. 2013.
{Finding Extraterrestrial Life Using Ground-based High-dispersion
  Spectroscopy}.
\textit{\apj} 764:182

\bibitem[{Trinh} et~al.(2017){Trinh}, {Scholten}, {Bonardi}, {Buitink},
  {Corstanje} et~al.]{2017PhRvD..95h3004T}
{Trinh} TNG, {Scholten} O, {Bonardi} A, {Buitink} S, {Corstanje} A, et~al.
  2017.
{Thunderstorm electric fields probed by extensive air showers through their
  polarized radio emission}.
\textit{\prd} 95:083004

\bibitem[{Vorgul} \& {Helling}(2016)]{2016MNRAS.458.1041V}
{Vorgul} I, {Helling} C. 2016.
{Flash ionization signature in coherent cyclotron emission from brown dwarfs}.
\textit{\mnras} 458:1041--1056

\bibitem[{Woitke} \& {Helling}(2003)]{2003A&A...399..297W}
{Woitke} P, {Helling} C. 2003.
{Dust in brown dwarfs. II. The coupled problem of dust formation and
  sedimentation}.
\textit{\aap} 399:297--313

\bibitem[{Woitke} et~al.(2017){Woitke}, {Helling}, {Hunter}, {Millard},
  {Turner} et~al.]{2017arXiv171201010W}
{Woitke} P, {Helling} C, {Hunter} GH, {Millard} JD, {Turner} GE, et~al. 2017.
{Equilibrium chemistry down to 100 K - Impact of silicates and phyllosilicates
  on carbon/oxygen ratio}.
\textit{ArXiv e-prints}

\bibitem[{Wong} et~al.(2016){Wong}, {Knutson}, {Kataria}, {Lewis}, {Burrows}
  et~al.]{2016ApJ...823..122W}
{Wong} I, {Knutson} HA, {Kataria} T, {Lewis} NK, {Burrows} A, et~al. 2016.
{3.6 and 4.5 {$\mu$}m Spitzer Phase Curves of the Highly Irradiated Hot
  Jupiters WASP-19b and HAT-P-7b}.
\textit{\apj} 823:122

\bibitem[{Wong} et~al.(2015){Wong}, {Knutson}, {Lewis}, {Kataria}, {Burrows}
  et~al.]{2015ApJ...811..122W}
{Wong} I, {Knutson} HA, {Lewis} NK, {Kataria} T, {Burrows} A, et~al. 2015.
{3.6 and 4.5 {$\mu$}m Phase Curves of the Highly Irradiated Eccentric Hot
  Jupiter WASP-14b}.
\textit{\apj} 811:122

\bibitem[{Zellem} et~al.(2014){Zellem}, {Lewis}, {Knutson}, {Griffith},
  {Showman} et~al.]{2014ApJ...790...53Z}
{Zellem} RT, {Lewis} NK, {Knutson} HA, {Griffith} CA, {Showman} AP, et~al.
  2014.
{The 4.5 {$\mu$}m Full-orbit Phase Curve of the Hot Jupiter HD 209458b}.
\textit{\apj} 790:53

\bibitem[{Zhang} \& {Showman}(2018)]{2018arXiv180309149Z}
{Zhang} X, {Showman} AP. 2018.
{Global-Mean Vertical Tracer Mixing in Planetary Atmospheres}.
\textit{ArXiv e-prints}

\end{thebibliography}

\end{document}